\documentclass[%
aps,
pre,
superscriptaddress,
showpacs,showkeys,
a4paper,
12pt,
longbibliography,
reprint,
notitlepage
]{revtex4-1}
\usepackage[english]{babel}
\usepackage{amssymb,amsmath,stmaryrd,array}

\usepackage{graphicx}
\usepackage{subfig}

\makeatletter

\newcommand{\ket}[1]{\ensuremath{|{#1}\rangle}}
\newcommand{\bra}[1]{\ensuremath{\langle{#1}|}}
\newcommand{\braket}[2]{\ensuremath{\langle{#1}|{#2}\rangle}}
\newcommand{\sca}[2]{\ensuremath{\bigl({#1}\cdot{#2}\bigr)}}
\newcommand{\avr}[1]{\ensuremath{\langle{#1}\rangle}}

\newcommand{\cnj}[1]{{#1}^{\ast}}
\newcommand{\hcnj}[1]{{#1}^{\dagger}}

\newcommand{\prt}[1]{\partial_{#1}}

\newcommand{\diag}{\mathop{\rm diag}\nolimits}

\renewcommand{\Re}{\mathop{\rm Re}\nolimits}
\renewcommand{\Im}{\mathop{\rm Im}\nolimits}
\newcommand{\Tr}{\mathop{\rm Tr}\nolimits}

\newcommand{\mum}{$\mu$m}
\newcommand{\dega}{\ensuremath{^\circ}}

 \newcommand{\bs}[1]{\boldsymbol{#1}}
 \newcommand{\vc}[1]{\mathbf{#1}}
 \newcommand{\mvc}[1]{\mathbf{#1}}
 \newcommand{\uvc}[1]{\hat{\mathbf{#1}}}
 
 \newcommand{\ind}[1]{\mathrm{#1}}

\newcommand{\dd}{\mathrm{d}}

 \newcommand{\e}{\mathrm{e}}

\makeatother

\begin{document}
\DeclareGraphicsExtensions{.jpg,.pdf,.eps,.png}

\title{Interferometric and Uhlmann phases of mixed polarization
  states}

\author{Alexei~D.~Kiselev}
\email[Email address: ]{alexei.d.kiselev@gmail.com}
\affiliation{%
 Saint Petersburg National Research University of Information Technologies,
 Mechanics and Optics (ITMO University),
 Kronverksky Prospekt 49,
 197101 Saint Petersburg, Russia}

\author{Vladimir~V.~Kesaev}
 \email[Email address: ]{vladimir.kesaev@gmail.com}
\affiliation{%
Lebedev Physical Institute,
Leninsky Prospekt 53, 119991 Moscow, Russia
 }

\date{\today}

\begin{abstract}
In our investigation into
the effects of the degree of polarization
in modulation of partially polarized light
we assume general settings
of the interferometry of partially polarized lightwaves
and perform theoretical analysis
of the geometric phases:
the Uhlmann phase and the interferometric
phase.
We introduce the relative Uhlmann
phase 
determined by the Uhlmann holonomies
of interfering beams
and show
that the interferometric phase
generalized to the case of nonunitary evolution
can, similar to the Uhlmann phase, 
be cast into the holonomy defined form.
By using the technique based on a two-arm
Mach-Zehnder interferometer,
two different dynamical regimes of light modulation
are experimentally studied:
(a)~modulation of the input light by the rotating quarter-wave plate (QWP);
and
(b)~modulation of the testing beam
by a birefringent plate
with electrically controlled anisotropy
represented by
the deformed-helix ferroelectric liquid crystal (DHFLC) cell.
In the setup with the rotating QWP,
the interferometric phase is found to be equal
to the relative Uhlmann phase.
Experimental and theoretical results
being in excellent agreement
both show that this phase is an oscillating function of
the QWP angle
and increases with the degree of polarization.
For modulation by the DHFLC cell,
the data
derived from our electro-optic measurements
are fitted using the theory of the orientational Kerr effect in FLCs.
This theory in combination with
the results of fitting is used to evaluate electric field dependencies
of the interferometric and the Uhlmann phases.
\end{abstract}

 \maketitle

%%%%%%%%%%%%%%
\section{Introduction}
\label{sec:intro}
%%%%%%%%%%%%%%

The notion of the phase of a
lightwave field
is essential to the understanding
of all interference
and diffraction phenomena in optics
where it plays a central role
as one of the determining factors
that, in particular, govern
certain topological properties of waves.
Topological aspects related to  
phases of optical fields
are among 
the most fascinating and intensely studied
subjects that 
have a long history dating back to
the original paper by 
Pancharatnam~\cite{Pancharatnam:proc:1956}
(see also a collection of important papers~\cite{Shapere:bk:1989}
and reviews~\cite{Anandan:nature:1992,Bhandari:physrep:1997,Hariharan:progr_opt:2005}). 
The Pancharatnam phase can naturally be defined as 
the phase acquired  by a light wave
as it evolves along a path in
the space of polarization states.
As it was originally discussed in Ref.~\cite{Berry:jmo:1987},
general concepts and approaches formulated
by Pancharatnam are closely related to 
the famous adiabatic quantal phase
(the Berry phase)
introduced by Berry in the well-known paper~\cite{Berry:prsla:1984}.

Berry~\cite{Berry:prsla:1984} 
analyzed the problem of a quantum mechanical
state developing adiabatically in time with a slowly varying
parameter dependent Hamiltonian. He has shown that when
the parameters return to their initial values after traversing a
closed path, the wavefunction acquires a ``geometric'' phase
factor, dependent on the path, in addition
to the well-known
``dynamical'' phase factor. Aharonov and
Anandan~\cite{Aharonov:prl:1987} 
removed the adiabatic restriction and replaced
the notion of parameter space by the notion of projective space
of rays in Hilbert space.
Samuel and Bhandari~\cite{Samuel:prl:1988}
extended the ideas of Pancharatnam to the cases
of non-cyclic and nonunitary evolutions.
A unifying kinematic approach 
that can be applied to these cases
was developed
by Mukunda and Simon~\cite{Mukunda:annphys:1993}.
The kinematic method was recently applied to
analyze geometric phases
associated with polarizing processes
of a monochromatic light wave~\cite{Lages:physe:2014}.
The geometric phase for non-cyclic
polarization changes was also studied 
experimentally in~\cite{vanDijk:optexp:2010}. 

In optics, the
geometric phase is also known as 
the Pancharatnam-Berry (PB) phase 
and can roughly be regarded as
a phase retardation that is exclusively 
determined by the geometry of transformations 
imposed on light by the medium. 
Optical devices~---~the so-called PB optical elements~---~
exploit the medium anisotropy
to introduce spatially dependent modulation of 
the polarization state of light across 
the plane transverse to propagation. 
In particular, such modulation results in a spatially inhomogeneous 
PB phase giving rise to a reshaped optical wavefront~\cite{Jisha:pra:2017}.
PB optical elements 
has
already been implemented
using various
architectures such as patterned subwavelength
gratings~\cite{Bomzon:optl:2:2002}, 
liquid crystals~\cite{Marrucci:apl:2006}
and metasurfaces~\cite{Capasso:nature:2014}. 

A problem
which is of key importance 
for both fundamental and technological reasons
concerns an extension of
the well-established results for the geometric phases
of pure states to a more general
case of mixed states.
Such states are described by mixed state density matrices
characterizing open physical systems that are not
perfectly isolated from their environment.
For optical wavefields,
mixed polarization states
of partially polarized waves
provide an important well-known example. 

Geometric phase 
for mixed states 
generalized to the realm
of density matrices
was first
introduced by
Uhlmann~\cite{Uhlmann:repmathph:1986,Uhlmann:annphys:1989}
(a review of the results of Uhlmann's approach
can also be found in~\cite{Uhlmann:repmathph:1995}).
Uhlmann's method deals with the gauge structure
arising from redundancy in 
the representation of density matrices
by pure (purified) states in a larger system
extended with additional ancilla degrees of freedom. 
Such a representation
of a density matrix $\rho$
is known as the purification
and
is mathematically equivalent
to putting the matrix into the factorized form:
$\rho=\vc{w}\hcnj{\vc{w}}$,
so that the purifications are defined 
as the Hilbert-Shmidt operators
$\vc{w}$ (the so-called amplitudes). 

The \textit{Uhlmann geometric phase} 
being the subject of numerous 
mathematical studies~\cite{Hubner:pla:1992,Hubner:pla:1993,Dittman:jgp:1992,Dabrowski:lmp:1990,
Ericsson:prl:2003,Budich:pra:2015}
has been extensively used
as a theoretical tool to characterize thermal topological order
in a variety of condensed matter systems such as
topological insulators and 
topological 
superconductors~\cite{Viyuela:prl:1:2014,Viyuela:prl:2:2014,Zhoushen:prl:2014,
Viyuela:2dmat:2015,Bruno:prl:2017,Grudst:prb:2017}. 
Despite some progress~\cite{Zhu:epl:2011,Viyuela:npj:2018},
reliable measurements of the Uhlmann phase
still remain a challenging problem.
As far as photonics is concerned,
this phase has not yet been explored in any detail.

In contrast to mathematically motivated concepts
behind the Uhlmann phase,
an alternative approach
to geometric phases of mixed states
put forward in Ref.~\cite{Sjoqvist:prl:2000}
is from outset built upon 
the interferometry based construction
leading to the so-called \textit{interferometric phase}.
Measurements of this phase
were performed
using 
a variety of experimental techniques such as
the single photon interferometry~\cite{Ericsson:prl:2005},
the NMR technique~\cite{Du:prl:2003,Ghosh:pla:2013,Zhu:epl:2011}
and the polarimetric method~\cite{Barberena:pra:2016}. 
As opposed to the Uhlmann phase,
the above interferometric approach requires 
additional analysis when applied to
physical systems with dissipative (nonunitary) 
dynamics~\cite{Tong:prl:2004,Yin:pra:2009}.

In this paper,
the Uhlmann and interferometric phases of mixed polarization states
within general settings of the interferometry of partially polarized
light fields will be our primary concern.
 Our theoretical considerations leading
to the relative geometric phases
behind the interference patterns 
provide the results in a unified form
applicable to both unitary (lossless) and 
nonunitary (dissipative) dynamics.
Specifically, we introduce the relative Uhlmann phase
for interfering beams
and show how the relative interferometric phase
can be generalized to the case of nonunitary
dynamics.

In our experiments 
based on the Mach-Zehnder
two-arm interferometer,
we investigate into the effects of 
the degree of polarization
in modulation of partially polarized light.
We study two different 
dynamical regimes of modulation
which are
governed by a rotating quarter-wave plate
(QWP)
and an electrically driven
ferroelectric liquid crystal cell.
Theoretical results applied to 
interpret the experimental data,
in particular, show that,
by contrast to the liquid crystal modulator,
for the rotating QWP,
the relative Uhlmann phase
appears to be coincident with
the 
experimentally measured interferometric phase.

The paper is organized as follows.
In Sec.~\ref{sec:modul-part-polar}
we 
introduce necessary notations
along with algebraic relations
and outline 
interferometry based considerations
that underline
the Pancharatnam phase
and the Pancharatnam function.

In Sec.~\ref{subsec:amplituted} 
we discuss how
to define the relative Uhlmann phase
for partially polarized waves
that acquire different Uhlmann phases
in the course of their evolution.
We find that such a gauge invariant phase
requires additional assumptions  to
make it uniquely determined.
This is demonstrated by introducing
two different relative Uhlmann phases.

An alternative line of reasoning leading to
the interferometric phase is presented 
in Sec.~\ref{subsec:interferometric}.
Our analysis will show how to deal with difficulties
that arise from  the complicated structure
underlying nonunitary evolution.
We find that the modified expression for the
interferometric phase
can be written in the form similar to
the Uhlmann phase
and 
can be regarded as a generalization of the phase 
to the case of nonunitary evolution.

The key common elements
of our experimental setup 
are described in Sec.~\ref{subsec:setup}.
In Sec.~\ref{subsec:qwp}
we present 
the experiments, where
the partially polarized light
passes through
the rotating quarter-wave plate
before being divided
into two beams
(they will be referred to as the
reference and testing beams), along with the results
and theoretical analysis. 

In Sec.~\ref{subsec:dhflc}
we briefly
discuss the orientational Kerr
effect
in deformed-helix liquid crystal (DHFLC)
cells 
and describe
the experiments
with the partially polarized testing beam
propagating through the DHFLC cell.
The parameters extracted from fitting
the experimental data are used
to evaluate both the interferometric
and Uhlmann phases.

Finally, in Sec.~\ref{sec:conclusion}
we draw the results together and make some concluding
remarks.
Mathematical details on computing
the Uhlmann connection
and the geometric
phases for absorbing (nonunitary)
DHFLC cells
are relegated to
Appendices~\ref{sec:appendix}
and~\ref{sec:appendix2},
respectively.

% %%%%%%%%%%%%%%
\section{Theory}
\label{sec:theory}
% %%%%%%%%%%%%%%

%%%%%%%%%%%%%%%%%%%%
\subsection{Modulation of partially polarized light
and Pancharatnam phase}
\label{sec:modul-part-polar}
%%%%%%%%%%%%%%%%%%%%

In this subsection, we introduce notations
and briefly discuss
the general theoretical structure
describing
interferometric measurements.
 
In a typical experimental setup
(see, e.g., Fig.~\ref{fig:mz_qwp}) 
based on a Mach-Zehnder two-beam interferometer,
a beam splitter  divides a collimated laser light
into two beams.
These beams~---~the so-called
\textit{reference} and \textit{testing (sample)} 
beams~---~evolve propagating in the corresponding arm
and, after reflection at the mirrors, 
are recombined at the semireflecting surface
of  the beam splitter. 
Then the interfering beams
emerging from the interferometer 
are projected by the lens 
on to a screen with a pinhole. 

In such a setup,
the output (total) beam
represented by
the vector amplitude, $\vc{E}$,
is a sum of two waves:
$\mathbf{E}=\mathbf{E}_1+\mathbf{E}_2$.
We shall use
the vectors of circular basis
$\uvc{e}_{\pm}=(\uvc{x}\pm i\uvc{y})/\sqrt{2}$
assuming that the vector amplitudes
of the waves, 
$\vc{E}_1$ and $\vc{E}_2$,
are defined by the circular components,
$E_{\pm}^{(1)}$ and
$E_{\pm}^{(2)}$, respectively.
It will also be convenient to use the bra-ket notations
for the amplitudes and their inner product:
\begin{align}
&
  \label{eq:E_total} 
\mathbf{E}_i
=
  \begin{pmatrix}
    E_{+}^{(i)}\\
    E_{-}^{(i)}
  \end{pmatrix}
\equiv \ket{\psi_i}
=\sum_{\mu=\pm 1}
E_{\mu}^{(i)}\ket{\mu},
\notag
\\
&
\cnj{\vc{E}}_{i}\cdot\vc{E}_{j}=
\braket{\psi_i}{\psi_j}
\end{align}
where
an asterisk will indicate complex conjugation
and $\ket{\pm}$ denote the vectors of
the circular orthonormal basis $\uvc{e}_{\pm}$
that meets the orthogonality conditions:
$\braket{\mu}{\nu}=\delta_{\mu\nu}$
($\delta_{\mu\nu}$ is the Kronecker delta).

We assume that the waves are 
linearly related to the input beam
$\vc{E}_0\equiv\ket{\psi_0}$
through the transmission 
matrices,
$\vc{T}_1$ and $\vc{T}_2$, 
as follows
\begin{align}
  \label{eq:lin_rel}
  \ket{\psi_1}=\e^{i\Phi_0}\vc{T}_1\ket{\psi_0},
\quad
  \ket{\psi_2}=\vc{T}_2\ket{\psi_0},
\end{align}
where
the relative phase $\Phi_0$~---~the 
so-called $\vc{U}(1)$ phase~---~ can be observed
in the output signal of the interferometer
producing the interference oscillations
(the interference fringes) as the phase varies.

An important point is that
the transmission matrices depend on the governing parameter, 
$0\le\tau\le s$,
and determine
the dynamics (evolution) of the wavestates:
$\ket{\psi_0}\mapsto \vc{T}_i(\tau)\ket{\psi_0}\equiv\ket{\psi_i(\tau)}$.
These matrices will generally be regarded as 
the evolution operators
with the waves exiting 
the interferometer taken at the final point $\tau=s$.
When the dynamics is lossless,
the operators $\vc{T}_i$ are unitary:
$\hcnj{\vc{T}}_i\vc{T}_i=\vc{I}$,
where the dagger denotes Hermitian
conjugation and $\vc{I}$
is the identity operator (matrix).
This is the case known as the unitary evolution.
Another important case occurs when
the wave $\ket{\psi_1}$ representing
the reference beam is characterized by
the transmission matrix $\vc{T}_1(\tau)=\vc{T}_1(0)$
which is independent of the governing parameter.

We consider
the general case of 
a  partially polarized input beam
with the degree of polarization equal to
$P_0$ that undergoes either unitary (lossless) or nonunitary 
(dissipative) evolution
as the governing parameter, $\tau$, varies from
$\tau=0$ to $\tau=s$.
This beam is characterized by the $2\times 2$
equal-time coherence
matrix~\cite{Mandl:bk:1995,Brosseau:bk:1998}:
\begin{align}
  \label{eq:M_0}
  \mathbf{M}_0=\bigl\langle
\mathbf{E}_0\otimes
\mathbf{E}_0^{\ast}
\bigr\rangle_t=I_0 \rho_0,
\end{align}
where
$\otimes$
and $\avr{...}_t$
indicate the tensor (dyadic) product
and the time averaging, respectively;
$I_0=\langle\mathbf{E}_0\cdot\mathbf{E}_0^{\ast}\rangle
=\mathrm{Tr}\mathbf{M}_0$
is the intensity of the incident wave
and
$\rho_0$
is
the normalized coherency matrix
($\mathrm{Tr}\rho_0=1$).
This matrix 
is known to play the role of
the density matrix 
describing
the mixed polarization state
and will be referred to as
\textit{the polarization density matrix}.
It
can generally be expressed
in terms of the eigenstates
as follows
\begin{align}
  \label{eq:rho_0}
  \rho_0=
\sum_{\mu=\pm 1}
p_\mu(0)\ket{p_\mu^{(0)}}\bra{p_\mu^{(0)}},
\quad
p_\mu(0)=\frac{1+\mu P_0}{2},
\end{align}
where the eigenpolarization vectors,
$\ket{p_{+}^{(0)}}$ and $\ket{p_{-}^{(0)}}$,
form the orthonormal basis:
$\braket{p_{\mu}^{(0)}}{p_\nu^{(0)}}=\delta_{\mu\nu}$.
The Stokes vector of the mixed state~\eqref{eq:rho_0}
\begin{align}
  \label{eq:stokes-input}
  (S_1^{(0)},S_2^{(0)},S_3^{(0)})=
I_0
P_0\, \uvc{s}_{0},
\end{align}
is proportional to
the normalized unit Stokes vector
\begin{align}
  \label{eq:s0-1}
  \uvc{s}_0=
(\sin(2\theta_0)\cos(2\phi_0),\sin(2\theta_0)\sin(2\phi_0),\cos(2\theta_0))
\end{align}  
characterizing the polarized 
part of the input wave.

Now we recast the initial density matrix $\rho_0$
into the explicit matrix form:
\begin{align}
  \label{eq:rho_0-circ}
  2\rho_0=
\mvc{C}_0
[
\bs{\sigma}_0+
P_0 
\bs{\sigma}_3
]
\hcnj{\mvc{C}}_0=\bs{\sigma}_0+P_0\sca{\uvc{s}_0}{\bs{\sigma}},
\end{align}
where
$\bs{\sigma}_0=\diag(1,1)$,
$\sca{\uvc{s}_0}{\bs{\sigma}}\equiv
{s}_1^{(0)}\bs{\sigma}_1+
{s}_2^{(0)}\bs{\sigma}_2+
{s}_3^{(0)}\bs{\sigma}_3$
and
$\bs{\sigma}_i$ with $i\in\{1,2,3\}$
are the Pauli matrices given by
\begin{align}
  \label{eq:Pauli}
  \bs{\sigma}_1=
  \begin{pmatrix}
    0&1\\
1&0
  \end{pmatrix},
\:
  \bs{\sigma}_2=
  \begin{pmatrix}
    0&-i\\
i&0
  \end{pmatrix},
\:
  \bs{\sigma}_3=
  \begin{pmatrix}
    1&0\\
0&-1
  \end{pmatrix}.
\end{align}
The components of the vectors
$\ket{p_{+}^{(0)}}$ and $\ket{p_{-}^{(0)}}$
enter the columns of the matrix
of initial eigenpolarization vectors
$\vc{C}_0$.
The elements of $\vc{C}_0$
can be conveniently expressed in terms of
the Wigner $D$ functions~\cite{Biedenharn:bk:1981,Varshalovich:bk:1988}
with the one-half angular momentum, $J=1/2$:
$\vc{D}_{\mu/2,\nu/2}^{1/2}(2\alpha,2\beta,2\gamma)\equiv 
\vc{D}_{\mu\nu}(\alpha,\beta,\gamma)$,
so that
$\vc{C}^{(0)}_{\mu\nu}=\vc{C}_{\mu\nu}(\phi_0,\theta_0)=\braket{\mu}{p_{\nu}^{(0)}}=
\vc{D}_{\mu\nu}(\phi_0,\theta_0,0)$
\begin{align}
&
  \label{eq:C0}
  \mvc{C}(\phi_0,\theta_0)=
\mvc{E}(\phi_0)\mvc{R}(\theta_0),
\:
\mvc{E}(\phi_0)=
\begin{pmatrix}
  \e^{-i\phi_0} & 0\\
0 & \e^{i\phi_0}
\end{pmatrix},
\notag
\\
&
\mvc{R}(\theta_0)=\e^{-i\theta_0\bs{\sigma}_2}=
\begin{pmatrix}
    \cos\theta_0 & -\sin\theta_0\\
    \sin\theta_0 & \cos\theta_0
  \end{pmatrix},
\end{align}
where the angle $0\le 2\phi_0\le 2\pi$ 
($0\le 2\theta_0\le \pi$)
is the azimuthal (polar) angle of the normalized Stokes
vector~\eqref{eq:s0-1}.
The angles $\phi_0$ and $\theta_0$
define the polarization ellipse parameters
for the wave with the eigenpolarization vector
$\ket{p_{+}^{(0)}}$:
$\phi_0$ is the \textit{polarization azimuth}
and $\epsilon_{\mathrm{ell}}=\tan\chi $ is the
\textit{ellipticity}, where 
$-\pi/4\le \chi=\pi/4-\theta_0\le\pi/4$ is the \textit{ellipticity angle}.

Note that the right-hand side of Eq.~\eqref{eq:rho_0-circ}
is an immediate consequence of
the following algebraic identities for 
the rotation matrix of the Wigner $D$ functions~\eqref{eq:C0}
and the Pauli matrices:
\begin{align}
  \label{eq:rel-Pauli-1}
&
  \mvc{C}(\phi,\theta)\bs{\sigma}_i\hcnj{\mvc{C}}(\phi,\theta)=
\sca{\uvc{s}_i}{\bs{\sigma}},
\\
  \label{eq:rel-Pauli-2}
&
\hcnj{\mvc{C}}(\phi,\theta) 
\bs{\sigma}
% \begin{pmatrix}
% \bs{\sigma}_1\\
% \bs{\sigma}_2\\
% \bs{\sigma}_3  
% \end{pmatrix}
\mvc{C}(\phi,\theta)=
\uvc{s}_1\bs{\sigma}_1+
\uvc{s}_2\bs{\sigma}_2+
\uvc{s}_3\bs{\sigma}_3,
\end{align}
where 
$\sca{\uvc{s}_i}{\bs{\sigma}}\equiv
{s}_1^{(i)}\bs{\sigma}_1+
{s}_2^{(i)}\bs{\sigma}_2+
{s}_3^{(i)}\bs{\sigma}_3$
and the unit vectors
\begin{align}
  \label{eq:rel-s0-i}
&
\uvc{s}_1=(\cos(2\theta)\cos(2\phi),\cos(2\theta)\sin(2\phi),-\sin(2\theta)),
\notag
\\
&
\uvc{s}_2=(-\sin(2\phi),\cos(2\phi),0),
\\
&
\uvc{s}_3\equiv\uvc{s}=
(\sin(2\theta)\cos(2\phi),\sin(2\theta)\sin(2\phi),\cos(2\theta))
\notag
\end{align}
meet the orthogonality conditions:
$\sca{\uvc{s}_i}{\uvc{s}_j}=\delta_{ij}$. 
[A more general version of the above
identities is given by
Eqs.~\eqref{eq:rel-Pauli-D1}--\eqref{eq:rel-S}
of Appendix~\ref{sec:appendix}.]

The density matrix of the output
wavefield 
(the beam emerging from the
interferometer)
at $\tau=s$
is given by
\begin{align}
&
  \label{eq:M-out}
  \rho=
\rho_1(s)+\rho_2(s)+
\e^{-i\Phi_{0}} \rho_{12}(s)+\e^{i\Phi_{0}} \rho_{21}(s),
\\
&
\label{eq:M_i-1}
\rho_{ij}(s)=
\hcnj{\rho}_{ji}(s)=
\vc{T}_j(s) \rho_0\hcnj{\vc{T}}_i(s),
\:
\rho_{i}(s)=\rho_{ii}(s).
\end{align}
The total intensity of the beams exiting the interferometer
\begin{align}
  \label{eq:I-tot}
  I=\langle\mathbf{E}\cdot\mathbf{E}^{\ast}\rangle=I_0 \mathrm{Tr}\mathbf{\rho}
=I_1+I_2+I_{12}
\end{align}
gives the intensity of the
interference pattern
and,
in addition to the contributions coming from
the density matrices of
the reference and testing beams  
($I_1=I_0 \mathrm{Tr}\mathbf{\rho}_1$ and $I_2=I_0 \mathrm{Tr}\mathbf{\rho}_2$, respectively),
contains
the interference part $I_{12}$,
determined by the interference terms of 
the density 
matrix~\eqref{eq:M_i-1}.
The latter is the only part of the intensity
that depends on the phase shift $\Phi_0$.
This part can be written in the form:
\begin{align}
  \label{eq:interference-intensity}
  I_{12}/I_0= 2\Re [\e^{-i\Phi_0} F_P]=2V\cos(\Phi_P-\Phi_0),
\end{align}
where $F_P$ is
the averaged product of transmission matrices 
$\vc{T}=\hcnj{\vc{T}}_1\vc{T}_2$
which is given by
\begin{align}
&
  \label{eq:PhiP}
F_P=\mathrm{Tr}[\vc{T}_2(s) \rho_0\hcnj{\vc{T}}_1(s)]=
V\exp(i\Phi_P),
\notag
\\
&
\Phi_P=\arg F_P.
\end{align}
Formula~\eqref{eq:interference-intensity}
introduces 
the total relative \textit{Pancharatnam phase},
$\Phi_P$, and
the \textit{visibility} of the interference pattern, 
$V$, through the averaged transition matrix
$\vc{T}=\hcnj{\vc{T}}_1(s)\vc{T}_2(s)$
that will be referred to as the \textit{Pancharatnam function}.

\subsection{Amplitudes and Uhlmann phase}
\label{subsec:amplituted}

The geometric phases 
of the interfering beams
generally
determine the part of  the Pancharatnam
phase~\eqref{eq:PhiP}
which is solely dictated
by the geometry
of the paths:
$\tau\mapsto\rho_i(\tau)$
with $0\le\tau\le s$.
These paths are
governed by
the dynamics of the density
matrices:
$\rho_1(\tau)=\vc{T}_1(\tau) \rho_0\hcnj{\vc{T}}_1(\tau)$
and $\rho_2(\tau)=\vc{T}_2(\tau) \rho_0\hcnj{\vc{T}}_2(\tau)$,
and represent the trajectories in the space of density matrices.
In this subsection, we discuss the general approach
to geometric phases of mixed states
originally developed by
Uhlmann in
Refs.~\cite{Uhlmann:repmphys:1986,Uhlmann:repmathph:1986,Uhlmann:annphys:1989,Uhlmann:repmathph:1995}.
This approach naturally leads to 
the geometric phase~---~the so-called \textit{Uhlmann phase}~---~
that generalizes the PB phase to the case of mixed polarization
states. 
 This phase is rarely discussed in the context
of optics and we briefly outline 
the basics of Uhlmann's approach
presenting the results
in the form suitable for our subsequent considerations.
Our primary goal is to define the relative Uhlmann phase
for interfering beams.

First we show that
the Pancharatnam phase can be
reexpressed in terms of the so-called \textit{amplitudes}
(the Hilbert-Schmidt operators)
$\vc{w}_i(\tau)$ 
that represent
the density matrices $\rho_i(\tau)$
in the ``squared amplitude'' form:
\begin{align}
  \label{eq:ampl_rho}
  \vc{w}_i(\tau)\hcnj{\vc{w}}_i(\tau)=\rho_i(\tau)\equiv \vc{T}_i(\tau) \rho_0\hcnj{\vc{T}}_i(\tau).
\end{align}
Alternatively, these amplitudes can be viewed as \textit{purifications}
where the density matrix is represented as a pure (purified) state
of a larger system. This extended system has 
additional degrees of freedom describing an ancilla
and the density matrix is obtained by
averaging the purified state over the ancilla.

The Hilbert-Schmidt space
of the operator amplitudes is 
equipped with the inner product 
\begin{align}
  \label{eq:inner-ampl}
  \avr{\vc{w},\vc{v}}=\Tr[\hcnj{\vc{w}}\vc{v}].
\end{align}
The Pancharatnam function~\eqref{eq:PhiP}
then can be cast into the form
of the inner product as follows
\begin{align}
  \label{eq:FP_inner}
  F_P=\Tr\bigl[
\hcnj{[\vc{T}_1(s)\sqrt{\rho_0}]}\vc{T}_2(s)\sqrt{\rho_0}
\bigr]
=\avr{\vc{w}_1(s),\vc{w}_2(s)}
\end{align}
and the relation
 \begin{align}
  \label{eq:PPh_inner}
  \Phi_P
=\arg \avr{\vc{w}_1(s),\vc{w}_2(s)}
\end{align}
with
\begin{align}
  \label{eq:ampl}
  \vc{w}_i(s)=\vc{T}_i(s)\sqrt{\rho_0}
\end{align}
gives
the total relative Pancharatnam phase
expressed in terms of the inner product
of the amplitudes~\eqref{eq:ampl}
characterizing the density matrices
of the interfering waves:
$\vc{w}_i(s)\hcnj{\vc{w}}_i(s)=\rho_i(s)$.
For a pure state, $\sqrt{\rho_0}$
is the projector $\ket{\psi}\bra{\psi}$
and 
this relation gives
the well-known Pancharatnam phase
 defined through
the inner product of the vector amplitudes:
$\braket{\vc{T}_1\psi}{\vc{T}_2\psi}$.

An important point is that
the amplitudes introduced by
the relation~\eqref{eq:ampl_rho} 
is not uniquely
defined and the gauge transformation:
\begin{align}
  \label{eq:GU}
  G_U:\:\vc{w}\mapsto \vc{w}_g=\vc{w} \vc{V}_g,
\end{align}
where $\vc{V}_g$ is a unitary operator,
leaves the density matrix unchanged: 
$\vc{w}\hcnj{\vc{w}}=\vc{w}_g\hcnj{\vc{w}_g}=\rho$.
In particular, this implies that any amplitude
$\vc{w}(\tau)$ can be written in the form:
$\sqrt{\rho(\tau)} \vc{V}(\tau)$.

In the geometry based Uhlmann 
approach~\cite{Uhlmann:repmphys:1986,Uhlmann:repmathph:1986,Uhlmann:annphys:1989,Uhlmann:repmathph:1995},
the amplitude $\vc{w}(\tau)$ is parallel transported
when it moves along the path with the shortest possible
length defined by the inner product~\eqref{eq:inner-ampl}:
$L=\int_0^{s}\sqrt{\avr{\prt{\tau}\vc{w}(\tau),\prt{\tau}\vc{w}(\tau)}}\dd\tau$.
It can be shown that this requires the tangent vector
$\prt{\tau}\vc{w}$ to be horizontal 
and meet the condition
\begin{align}
  \label{eq:PTU}
 \hcnj{\vc{w}} \prt{\tau}\vc{w}=[\prt{\tau}\hcnj{\vc{w}}]\vc{w},
\end{align}
where $\prt{\tau}$ stands for the derivative with respect to $\tau$.
This condition called 
the \textit{Uhlmann parallel transport} (UPT)
condition can be used to
find the \textit{parallel transported amplitudes}.
The procedure in its standard form
assumes
using the UPT condition for 
the amplitudes
$\tilde{\vc{w}}=\sqrt{\rho(\tau)} \vc{V}(\tau)$
to yield the unitary (the so-called \textit{Uhlmann holonomy}),
$\tilde{\vc{V}}_U$,
and obtain the parallel
transported amplitude: 
$\tilde{\vc{w}}_U=\sqrt{\rho(\tau)} \tilde{\vc{V}}_U(\tau)$.
This amplitude then gives the gauge
invariant phase known as the \textit{Uhlmann phase}:
$\Phi_U=\arg F_U$,
where $F_U=\Tr[\sqrt{\rho(0)}\sqrt{\rho(s)}\tilde{\vc{V}}_U(s)]$
is the \textit{Uhlmann function}.
Note that the above terminology originates from
differential geometry of fiber bundles
that provides natural mathematical structures
for understanding geometric 
phases~\cite{Simon:prl:1983,Bohm:pra:1991,BenAryeh:joptb:2004}.

For our purposes,
it is more convenient
to deal with the amplitudes
of the form:
$\vc{w}=\vc{T}(\tau)\sqrt{\rho(\tau)} \vc{V}(\tau)$.
For these amplitudes,
we briefly discuss how to find
 the Uhlmann holonomy
$\vc{V}_U$ and the parallel transported amplitude
\begin{align}
  \label{eq:w_U}
  \vc{w}_U(\tau)=\vc{T}(\tau)\sqrt{\rho_0}\,\vc{V}_U(\tau).
\end{align}

First we define the operator
\begin{align}
  \label{eq:H_U}
  [\prt{\tau}\vc{V}_U]\hcnj{\vc{V}}_U=i\vc{H}_{U},
\:
%\vc{T}^{-1}\prt{\tau}\vc{T} =i\vc{H},
\end{align}
which is Hermitian: 
$\hcnj{\vc{H}_U}=\vc{H}_U$
and
is known as the \textit{Uhlmann connection},
and substitute Eq.~\eqref{eq:w_U}
into the UPT condition~\eqref{eq:PTU}
to deduce the equation for $\vc{H}_U$
in the form:
\begin{align}
  \label{eq:H_U-1}
&
\sqrt{\rho_0}
\{
\hcnj{\vc{T}}\prt{\tau}\vc{T}-[\prt{\tau}\hcnj{\vc{T}}]\vc{T}
\}
\sqrt{\rho_0}=-i[\vc{H}_U,\tilde{\rho}]_{+},
% \sqrt{\rho_0}
% [\vc{A}\vc{H}+\hcnj{\vc{H}}\vc{A}
% ]
% \sqrt{\rho_0}=-[\vc{H}_U,\tilde{\rho}]_{+},
\\
&
\label{eq:TT}
\tilde{\rho}=\sqrt{\rho_0}\vc{A}\sqrt{\rho_0},
\quad
\vc{A}=\hcnj{\vc{T}}\vc{T},  
\end{align}
where $[A,B]_{+}=AB+BA$ stands for anticommutator.
% Another form of this equation 
% \begin{align}
%   \label{eq:H_U-2}
% &
  
% \end{align}
% can be a bit more convenient for calculations.

Given the Uhlmann connection,
the Uhlmann holonomy can be computed
by solving the initial value problem
\begin{subequations}
\label{eq:H_U_inival}
\begin{align}
&
  \label{eq:H_U_eq}
  \prt{\tau}\vc{V}_U=i\vc{H}_{U}\vc{V}_U,
\\
&
  \label{eq:H_U_ini}
\vc{V}_U(0)=\vc{I}.
%\quad \vc{T}(0)\sqrt{\rho_0}=\sqrt{\rho(0)}\vc{V}(0),
\end{align}  
\end{subequations}
% where $\rho(0)=\vc{T}(0)\rho_0\hcnj{\vc{T}}(0)$
% and the initial condition $\vc{w}(0)=\sqrt{\rho(0)}$
% gives the corresponding condition for
% $\vc{V}_U(0)$.
The Uhlmann phase
for the amplitude~\eqref{eq:w_U}
at $\tau=s$
can now be evaluated
as follows
\begin{align}
  \label{eq:PhiU_w}
&
\Phi_U(s)=\arg F_U(s),\quad
F_U(s)= \avr{\vc{w}_U(0),\vc{w}_U(s)},
\\
&
  \label{eq:FU_w}
F_U(s)=\Tr[
\sqrt{\rho_0}\hcnj{\vc{T}}(0)\vc{T}(s)\sqrt{\rho_0}\vc{V}_U(s)
]=
\notag
\\
&
\Tr[
\sqrt{\rho(0)}\sqrt{\rho(s)}\tilde{\vc{V}}_U
],
\:
\tilde{\vc{V}}_U=\hcnj{\vc{V}}(s)\vc{V}_U(s)\vc{V}(0),
\\
&
  \label{eq:V_w}
\vc{T}(s)\sqrt{\rho_0}\vc{V}(s)=\sqrt{\rho(s)}.
\end{align}
This phase defined as the phase
between the initial amplitude, $\vc{w}(0)=\vc{w}_U(0)$,
and the parallel-transported amplitude
$\vc{w}_U(s)$ [see Eq.~\eqref{eq:w_U}]
is gauge invariant.
More specifically, gauge invariance
means that, when the amplitude
$\vc{w}$ is changed
to $\vc{w}_g=\vc{w}\vc{V}_g$,
the Uhlmann connection
transforms into the operator:
$\vc{H}_U^{(g)}=
\hcnj{\vc{V}}_g\vc{H}_U\vc{V}_g+
i\hcnj{\vc{V}}_g\prt{\tau}\vc{V}_g$
giving the transformed holonomy:
$\vc{V}_U^{(g)}=
\hcnj{\vc{V}}_g\vc{V}_U\vc{V}_g(0)$,
so that the operators
$\hcnj{\vc{w}}(0)\vc{w}(s)\vc{V}_U(s)$
and
$\vc{V}_g(0)[\hcnj{\vc{w}}_g(0)\vc{w}_g(s)\vc{V}_U^{(g)}(s)]\hcnj{\vc{V}}_g(0)$
are identical.
It immediately follows that
the Uhlmann function
$F_U$ and the Uhlmann phase
$\Phi_U$ are both gauge invariant.

Interestingly, the Uhlmann function $F_U$
can be written as the inner product
of the following form:
\begin{align}
\label{eq:F_U}
&
F_U= \avr{\tilde{\vc{w}}_1,\tilde{\vc{w}}_2},
\:
\tilde{\vc{w}}_1=\vc{T}(0)\sqrt{\rho_0}\hcnj{\vc{V}}_U(s)=
\vc{T}_{U}\sqrt{\rho_0},  
\end{align}
where $\tilde{\vc{w}}_2=\vc{T}(s)\sqrt{\rho_0}$.
This relation implies that the Uhlmann phase
can be measured as the Pancharatnam
phase~\eqref{eq:PPh_inner}
provided the evolution of the amplitude $\vc{w}_1$
is tuned to
be governed by the Uhlmann operator 
$\vc{T}_{U}=\vc{T}(0)[\rho_0]^{1/2}\hcnj{\vc{V}}_U\rho_0^{-1/2}$.
Since this operator is generally nonunitary,
such a measurement would require
specifically tailored
nonunitary dynamics.

We can now define the relative Uhlmann phase
between the two parallel-transported amplitudes,
$\vc{w}_U^{(1)}$ and $\vc{w}_U^{(2)}$:
\begin{align}
  \label{eq:w_i}
  \vc{w}_U^{(i)}(s)=\vc{T}_i(s)\sqrt{\rho_0}\,\vc{V}_U^{(i)}(s),
\end{align}
where $i\in\{1,2\}$.
A natural generalization
of the Uhlmann phase~\eqref{eq:PhiU_w}
can, 
similarly to the Pancharatnam phase~\eqref{eq:FP_inner},
be formulated in terms
of the inner product of the amplitudes,
 $\avr{\vc{w}_U^{(1)},\vc{w}_U^{(2)}}$, as follows
\begin{align}
  \label{eq:PhiU_rel}
&
\Phi_U^{(12)}(s)=\arg F_U^{(12)}(s)-\arg F_U^{(12)}(0),
\\
&
  \label{eq:FU_rel}
F_U^{(12)}(s)=\avr{\vc{w}_U^{(1)},\vc{w}_U^{(2)}}=
\notag
\\
&
\Tr[
\sqrt{\rho_0}\hcnj{\vc{T}}_1(s)\vc{T}_2(s)\sqrt{\rho_0}\vc{V}_U^{(12)}(s)
],
\\
&
  \label{eq:V_rel}
\vc{V}_U^{(12)}(s)=\vc{V}_U^{(2)}(s)\hcnj{[\vc{V}_U^{(1)}(s)]},
\end{align}
where $\arg F_U^{(12)}(0)=\Phi_P(0)$
is the initial value of the 
Pancharatnam phase~\eqref{eq:FP_inner}
which can be incorporated into
the phase shift $\Phi_0$.
Hence 
its contribution is substracted from
the Uhlmann phase.
It can also be readily seen that the phase~\eqref{eq:PhiU_rel}
is invariant under the gauge transformation:
$\vc{w}_i(s)\mapsto \vc{w}_i(s)\vc{V}_i^{(g)}(s)$,
where the unitary gauge operators,
$\vc{V}_1^{(g)}$ and $\vc{V}_2^{(g)}$, are required
to satisfy the constraint:
$\vc{V}_1^{(g)}(0)=\vc{V}_2^{(g)}(0)$ 
imposed to keep the initial
relative phase unchanged.

Note that there is an alternative
method to introduce
the relative Uhlmann phase.
In this method, we use
a different form of the parallel-transported
amplitudes:
$\tilde{\vc{w}}_i(s)=\sqrt{\rho_i(s)}\tilde{\vc{V}}_U^{(i)}(s)$,
with 
the gauge transformation:
$\tilde{\vc{V}}_U^{(i)}(s)=\hcnj{\vc{V}}_i(s)\vc{V}_U^{(i)}(s)\vc{V}_i(0)$
[see Eq.~\eqref{eq:FU_w}]
linking the holonomies
$\tilde{\vc{V}}_U^{(i)}$ and $\vc{V}_U^{(i)}$, where
the relations
$\vc{T}_i(s)\sqrt{\rho_0}\vc{V}_i(s)=\sqrt{\rho_i(s)}$
[see Eq.~\eqref{eq:V_w}] define the unitaries
$\vc{V}_i(s)$.
Following the line of reasoning
presented above Eqs.~\eqref{eq:PhiU_w}--\eqref{eq:V_w},
we arrive at similar formulas given by
\begin{align}
  \label{eq:tPhiU_rel}
&
\tilde{\Phi}_U^{(12)}(s)=\arg \tilde{F}_U^{(12)}(s),
\\
&
  \label{eq:tFU_rel}
\tilde{F}_U^{(12)}(s)=\avr{\tilde{\vc{w}}_1(s),\tilde{\vc{w}}_2(s)}=
\notag
\\
&
\Tr\Bigl[
\sqrt{\rho_1(s)}\sqrt{\rho_2(s)}\tilde{\vc{V}}_U^{(2)}(s)\hcnj{[\tilde{\vc{V}}_U^{(1)}(s)]}
\Bigr]=
\notag
\\
&
\Tr\bigl[
\sqrt{\rho_0}\hcnj{\vc{T}}_1(s)\vc{T}_2(s)\sqrt{\rho_0}\tilde{\vc{V}}_U^{(12)}(s)
\bigr],
\\
&
  \label{eq:tV_rel}
\tilde{\vc{V}}_U^{(12)}(s)=\vc{V}_U^{(2)}(s)\vc{V}_{12}(0)\hcnj{[\vc{V}_U^{(1)}(s)]},
\end{align}
where
$\vc{V}_{12}(0)=\vc{V}_2(0)\hcnj{\vc{V}}_1(0)$.
The relative phases
$\tilde{\Phi}_U^{(12)}$ and $\Phi_U^{(12)}$
are both invariant under
the gauge transformations
with $\vc{V}_1^{(g)}(0)=\vc{V}_2^{(g)}(0)$.
In addition,
the phase $\tilde{\Phi}_U^{(12)}$
can be obtained from
$\Phi_U^{(12)}$ by applying
a gauge transformation
with $\vc{V}_i^{(g)}(0)=\vc{V}_i(0)$.
Such a transformation breaks
the condition $\vc{V}_1^{(g)}(0)=\vc{V}_2^{(g)}(0)$
giving the phase that differs
from $\Phi_U^{(12)}$.

%%%%%%%%%%%%%%%
\subsection{Interferometric phase}
\label{subsec:interferometric}
%%%%%%%%%%%%%%%

In this section,
we concentrate on the approach that was suggested
in Refs~\cite{Sjoqvist:prl:2000,Tong:prl:2004}.
This approach is based on an alternative
representation of the Pancharatnam
function~\eqref{eq:PhiP}
and gives the so-called \textit{interferometric phase},
$\Phi_I$.
We find that, 
for nonunitary dynamics,
this representation
has complicated structure
and 
use a generalized version of 
the interferometric parallel transport
(IPT)
conditions to
obtain the interferometric phase
expressed in terms of
the interferometric holonomies.
 
In order
to express the Pancharatnam function~\eqref{eq:PhiP}
in terms of eigenstates,
we begin with the singular value decomposition
(see, e.g., the textbook~\cite{Poole:bk:2015})
for the amplitudes~\eqref{eq:ampl}
given by
\begin{align}
  \label{eq:svd}
&
\vc{w}_i(s)=
  \vc{T}_i(s)\sqrt{\rho_0}=\sqrt{\rho_i(s)} \vc{V}_i(s)
\notag
\\
&
=\sum_{\mu}\sqrt{p^{(i)}_{\mu}(s)}\ket{p^{(i)}_{\mu}(s)}\bra{r^{(i)}_{\mu}(s)},
\end{align}
where
$\ket{p^{(i)}_{\mu}(s)}$ and $\ket{r^{(i)}_{\mu}(s)}$
are the normalized eigenvectors of 
the density matrix $\rho_i(s)$ and the Hermitian
operator $\tilde{\rho}_i(s)$, respectively:
\begin{subequations}
  \label{eq:eigen}
\begin{align}
\label{eq:lefteigen}
&
\rho_i(s)\ket{p^{(i)}_{\mu}(s)}=p^{(i)}_{\mu}(s)\ket{p^{(i)}_{\mu}(s)},
\\
&
\label{eq:eq:righteigen}
\tilde{\rho}_i(s)\ket{r^{(i)}_{\mu}(s)}=p^{(i)}_{\mu}(s) \ket{r^{(i)}_{\mu}(s)},
\\
&
  \label{eq:tilderho}
  \tilde{\rho}_i(s)=\hcnj{\vc{w}}_i(s)\vc{w}_i(s)=
\sqrt{\rho_0}\,\vc{A}_i(s)\sqrt{\rho_0},
\end{align}
\end{subequations}
where $\vc{A}_i(s)=\hcnj{\vc{T}}_i(s)\vc{T}_i(s)$.

By substituting the singular value decomposition
for the amplitudes given by Eq.~\eqref{eq:svd} 
into formula~\eqref{eq:PPh_inner}, 
we derive the representation
for the Pancharatnam phase:
\begin{align}
&
  \label{eq:P-phase-1}
  \Phi_P=\arg \sum_{\mu,\nu=\pm 1} F_{\mu\nu}^{(P)},
\quad F_{\mu\nu}^{(P)}=
\notag
\\
&
 \sqrt{p_{\mu}^{(1)}(s)p_{\nu}^{(2)}(s)}
\braket{p^{(1)}_{\mu}(s)}{p^{(2)}_{\nu}(s)}
\braket{r^{(2)}_{\nu}(s)}{r^{(1)}_{\mu}(s)},
\end{align}
where the Pancharatnam function
is written as a superposition of contributions
coming from the eigenstates:
$\ket{p^{(i)}_{\mu}(s)}$
and $\ket{r^{(i)}_{\mu}(s)}$.

As we have already discussed in the previous
subsection,
the geometric phase is a gauge invariant part of 
the Pancharatnam (total) phase
which is solely determined by
the path traced out by the polarization density matrix
in the space of mixed states.
This phase can generally be obtained 
using the parallel transport conditions to
eliminate the dynamical contributions.
For a pure state, $\ket{\psi}$, the standard
parallel transport prescription
requires
the tangent vector $\prt{\tau}\ket{\psi}$
to be horizontal leading
to the well-known parallel transport condition:
$\bra{\psi}\prt{\tau}\ket{\psi}=0$
that determines the dynamical phase.
 
The Pancharatnam function
expressed in terms of the eigenpolarization states
is the starting point of
the interferometry based
approach~\cite{Sjoqvist:prl:2000,Tong:prl:2004}.
This approach
deals with the gauge transformation
changing the phases of the eigenstates:
\begin{align}
  \label{eq:gauge-int}
 G_I:\:
& 
\ket{p^{(i)}_{\mu}}\mapsto
  \e^{i\alpha_{\mu}^{(i)}}\ket{p^{(i)}_{\mu}},
\notag
\\
&
  \ket{r^{(i)}_{\mu}}\mapsto
  \e^{i\beta_{\mu}^{(i)}}\ket{r^{(i)}_{\mu}}
\end{align}
and uses the PT condition for pure states
to obtain the dynamical phases. 

This strategy can be applied to the
generalized representation of the Pancharatnam phase~\eqref{eq:P-phase-1}.
We shall employ
a straightforward generalization of the IPT conditions
suggested in Refs~\cite{Sjoqvist:prl:2000,Tong:prl:2004}
which can be written as follows
\begin{align}
  \label{eq:PT_int}
\bra{r^{(i)}_{\mu}}\hcnj{\vc{V}_i} \prt{s}{\vc{V}_i}\ket{r^{(i)}_{\mu}}=0.
\end{align}
These conditions can now be combined with
formulas for $\vc{V}_i$
[see Eq.~\eqref{eq:svd}] 
to deduce the PT relations 
\begin{align}
  \label{eq:PT_int-1}
  \bra{p^{(i)}_{\mu}}\prt{s}\ket{p^{(i)}_{\mu}} - \bra{r^{(i)}_{\mu}}\prt{s}\ket{r^{(i)}_{\mu}}=0.
\end{align}
So, the transformation eliminating
the dynamical contributions reads
\begin{align}
  \label{eq:PT_transf}
  \ket{p^{(i)}_{\mu}}\mapsto
\e^{-i\phi^{(i)}_{\mu}}\ket{p^{(i)}_{\mu}},
\:
  \ket{r^{(i)}_{\mu}}\mapsto
\e^{-i\psi^{(i)}_{\mu}}\ket{r^{(i)}_{\mu}},
\end{align}
where the dynamical phases, 
$\phi^{(i)}_{\mu}$ and $\psi^{(i)}_{\mu}$,
are given by
\begin{subequations}
\label{eq:dyn_p-1}
\begin{align}
&
  \label{eq:dyn_phi-1}
  \phi^{(i)}_{\mu}=\phi^{(i)}_{\mu}(s)+\phi^{(i)}_{\mu}(0),
\notag
\\
&
-i \phi^{(i)}_{\mu}(s)=
  -\int_0^s\bra{p^{(i)}_{\mu}(\tau)}\prt{\tau}\ket{p^{(i)}_{\mu}(\tau)}\dd
  \tau
\\
\label{eq:dyn_psi-1}
&
 \psi^{(i)}_{\mu}=\psi^{(i)}_{\mu}(s)+\psi^{(i)}_{\mu}(0),
\notag
\\
&
  -i \psi^{(i)}_{\mu}(s)=
  -\int_0^s\bra{r^{(i)}_{\mu}(\tau)}\prt{\tau}\ket{r^{(i)}_{\mu}(\tau)}\dd
  \tau.
\end{align}
\end{subequations}
The parameter independent phases, $\phi^{(i)}_{\mu}(0)$
and $\psi^{(i)}_{\mu}(0)$,
are introduced to make the interferometric phase
\begin{align}
  \label{eq:Phi_I}
  \Phi_I=\arg \sum_{\mu,\nu=\pm 1} F_{\mu\nu}^{(I)},
\quad F_{\mu\nu}^{(I)}=F_{\mu\nu}^{(P)}\e^{-i\Phi_{\mu\nu}^{(d)}}
\end{align}
gauge invariant at $s=0$.
We can now write down the expressions for 
the dynamical phase that enters the geometric
phase~\eqref{eq:Phi_I}. 
The result is
\begin{align}
  \label{eq:dyn_int_fin}
  \Phi_{\mu\nu}^{(d)}=
\Phi_{\nu}^{(2)}-\Phi_{\mu}^{(1)},
\quad
\Phi_{\mu}^{(i)}=\phi^{(i)}_{\mu}-\psi^{(i)}_{\mu},
\end{align}
where
\begin{subequations}
\label{eq:ph0}
\begin{align}
&
  \label{eq:phi0}
\phi^{(2)}_{\nu}(0)-\phi^{(1)}_{\mu}(0)=
\arg \braket{p^{(1)}_{\mu}(0)}{p^{(2)}_{\nu}(0)},
\\
&
\label{eq:psi0}
\psi^{(2)}_{\nu}(0)-\psi^{(1)}_{\mu}(0)=
\arg \braket{r^{(1)}_{\mu}(0)}{r^{(2)}_{\nu}(0)}
\end{align}
\end{subequations}
and the dynamical phases, $\phi^{(i)}_{\mu}(s)$
and $\psi^{(i)}_{\mu}(s)$,
are defined in Eqs.~\eqref{eq:dyn_phi-1}
and~\eqref{eq:dyn_psi-1}, respectively.
Formulas~\eqref{eq:Phi_I}-~\eqref{eq:ph0}
give the resulting expression for the interferometric
phase that is
invariant under the transformations~\eqref{eq:gauge-int}
and $\Phi_I(0)=0$. 

In order to put the result into the form 
suitable for a comparison
with the Uhlmann phase, 
we define the unitary operators
\begin{align}
&
  \label{eq:tV_I}
  \tilde{\vc{V}}_{I}^{(i)}=\sum_{\mu=\pm
  1}\e^{-i\Phi_{\mu}^{(i)}}\ket{p^{(i)}_{\mu}(s)}\bra{p^{(i)}_{\mu}(s)},
\\
&
  \label{eq:V_I}
  \vc{V}_{I}^{(i)}=\sum_{\mu=\pm
  1}\e^{-i\Phi_{\mu}^{(i)}}\ket{r^{(i)}_{\mu}(s)}\bra{r^{(i)}_{\mu}(s)},
\end{align}
and notice that the transformation of the
eigenstates~\eqref{eq:PT_transf}
is equivalent to the transformation of the amplitudes:
\begin{align}
&
  \label{eq:tilde-w_i}
  \vc{w}_i(s)=
  \vc{T}_i(s)\sqrt{\rho_0}\mapsto
  \vc{w}_I^{(i)}(s)=\tilde{\vc{T}}_i(s)\sqrt{\rho_0}
\notag
\\
&
=\vc{w}_i(s)\vc{V}_I^{(i)}(s),
\quad
\tilde{\vc{T}}_i(s)=
\tilde{\vc{V}}_I^{(i)}(s)\vc{T}_i(s).
\end{align}
We can now express
the interferometric phase 
in terms of the transformed amplitudes as follows
\begin{align}
&
\label{eq:PhiI_rel}
\Phi_I(s)=\arg F_I^{(12)}(s),
\\
  &
  \label{eq:FI_rel}
F_I^{(12)}(s)=
\avr{\vc{w}_I^{(1)}(s),\vc{w}_I^{(2)}(s)}=
\Tr[
\rho_0\hcnj{\tilde{\vc{T}}}_1(s) \tilde{\vc{T}}_2(s)]
\notag
\\
&
=\Tr[
\sqrt{\rho_0}\,\hcnj{\vc{T}}_1(s)\vc{T}_2(s)\sqrt{\rho_0}\,\vc{V}_I^{(12)}(s)
],
\\
&
  \label{eq:VI_rel}
\vc{V}_I^{(12)}(s)=\vc{V}_I^{(2)}(s)\hcnj{[\vc{V}_I^{(1)}(s)]}.
\end{align}

Formulas~\eqref{eq:tV_I}--\eqref{eq:VI_rel}
combined with the expressions for the
dynamical phases~\eqref{eq:dyn_int_fin}--\eqref{eq:ph0}
is our key theoretical result for the interferometric
phase.

Interestingly, for mixed polarization states characterized 
by the regular initial density matrix,
the expressions for the dynamical phase
difference
$\phi_{\mu}^{(i)}(s)-\psi_{\mu}^{(i)}(s)$
[see Eqs.~\eqref{eq:dyn_phi-1}
and~\eqref{eq:dyn_psi-1}]
that enters the phase
$\Phi_{\mu}^{(i)}$
can be further simplified
using the relation
\begin{align}
  \label{eq:p_mu_r_mu}
  \hcnj{\vc{w}}_i(\tau)\ket{p_{\mu}^{(i)}(\tau)}=[p_{\mu}^{(i)}(\tau)]^{1/2}\ket{r_{\mu}^{(i)}(\tau)}
\end{align}
linking the eigenstates
of the density matrix, $\rho_i(\tau)=\vc{w}_i(\tau)\hcnj{\vc{w}}_i(\tau)$, and 
the operator:
$\tilde{\rho}_i(\tau)=\hcnj{\vc{w}}_i(\tau)\vc{w}_i(\tau)$.
After a rather straightforward algebra,
we have
\begin{align}
&
  \label{eq:pdotp-rdotr}
  -\bra{p^{(i)}_{\mu}(\tau)}\prt{\tau}\ket{p^{(i)}_{\mu}(\tau)}
+\bra{r^{(i)}_{\mu}(\tau)}\prt{\tau}\ket{r^{(i)}_{\mu}(\tau)}
\notag
\\
&
=i \Im
  \frac{\bra{p^{(i)}_{\mu}(\tau)}\vc{w}_i(\tau)\hcnj{\dot{\vc{w}}}_i(\tau)\ket{p^{(i)}_{\mu}(\tau)}}
{\bra{p^{(i)}_{\mu}(\tau)}\vc{w}_i(\tau)\hcnj{{\vc{w}}}_i(\tau)\ket{p^{(i)}_{\mu}(\tau)}}
\notag
\\
&
=
-i \Im \bra{p^{(i)}_{\mu}(\tau)}\dot{\vc{T}}_i(\tau)\vc{T}_i^{-1}(\tau)\ket{p^{(i)}_{\mu}(\tau)},
\end{align}
where a dot over the letter will indicate
the derivative with respect to $\tau$,
 so that the phase difference in the simplified form
is given by
 \begin{align}
   \label{eq:phi-psi}
   \phi_{\mu}^{(i)}(s)-\psi_{\mu}^{(i)}(s)=
\Im 
\int_{0}^{s}
\bra{p^{(i)}_{\mu}}\dot{\vc{T}}_i\vc{T}_i^{-1}\ket{p^{(i)}_{\mu}}\dd\tau.
 \end{align}

A comparison between Eqs.~\eqref{eq:PhiI_rel}--\eqref{eq:VI_rel}
and the expressions for the relative Uhlmann phases
[see Eqs.~\eqref{eq:PhiU_rel}--\eqref{eq:V_rel} and
Eqs.~\eqref{eq:tPhiU_rel}--~\eqref{eq:tV_rel}]
shows that, 
the operators~\eqref{eq:V_I}
play the role 
similar to the Uhlmann holonomies
and might be called the \textit{interferometric holonomies}.
The difference between these holonomies
is determined by the difference in the underlying gauge
structures and thus in the PT conditions.
It manifests itself
in quantitatively non-equivalent predictions for 
the geometric phases~\cite{Slater:lmathph:2002,Ericsson:prl:2003,Andersson:prsa:2016}.
Note that there are different unifying approaches to the geometric phases
put forward in Refs.~\cite{Rezakhani:pra:1:2006,Andersson:jpa:2015}.

Now we briefly comment on some special cases.
We begin with the case of unitary evolution,
where $\vc{T}_1$ and $\vc{T}_2$
are both unitary. 
In this case,
the operators 
$\tilde{\rho}_i(s)$
are equal to the initial
density matrix:
$\tilde{\rho}_i(s)=\rho_0$
and the eigenstates
$\ket{r_{\mu}^{(i)}}=\ket{p_{\mu}(0)}$
with $\psi_{\mu}^{(i)}=0$
are independent of the governing parameter.
So, the interferometric phase is
\begin{align}
  \label{eq:FI_uni}
  \Phi_I(s)=\arg \sum_{\mu=\pm 1}
p_{\mu}(0)
\braket{p^{(1)}_{\mu}(s)}{p^{(2)}_{\mu}(s)}
\e^{-i\Phi_{\mu}^{(d)}}
\end{align}
where
$\Phi_{\mu}^{(d)}=\phi_{\mu}^{(2)}-\phi_{\mu}^{(1)}$.
At $\vc{T}_1=\vc{I}$, 
the eigenstates
of the reference beam
are the eigenpolarization vectors of
the initial density matrix
$\rho_0=\rho_1(s)=\tilde{\rho}_1(s)$:
$\ket{p_{\mu}^{(1)}}=\ket{r_{\mu}^{(1)}}=\ket{p_{\mu}(0)}$
and formula~\eqref{eq:FI_uni} 
recovers  the interferometric
phase obtained in~Ref.~\cite{Sjoqvist:prl:2000}.

For nonunitary evolution with
$\vc{T}_1=\vc{I}$, our formulas [see Eq.~\eqref{eq:P-phase-1} and
Eq.~\eqref{eq:Phi_I}]
and the expressions given in Ref.~\cite{Tong:prl:2004}
are identical only if 
$\braket{r_\nu(s)}{p_{\mu}(0)}=\delta_{\mu\nu}$
(for the moment, the index indicating the testing beam is dropped).
The latter occurs when
$\rho_0$ and $\tilde{\rho}$ are commuting operators:
$[\tilde{\rho}(s),\rho_0]=0$,
so that
$\ket{r_\nu(s)}=\ket{p_\nu(0)}$.
But, for absorbing media,
the operator
$\hcnj{\vc{T}}(s)\vc{T}(s)$
and the density matrix
$\rho_0$
generally does not commute.

In conclusion, we consider the case of a pure state
with $P_0=1$ and $\rho_0=\sqrt{\rho_0}=\ket{\psi}\bra{\psi}$.
In this case, we have
$\ket{r_1}=\ket{r_2}=\ket{\psi}$
and the density matrices are given by
\begin{align}
&
  \label{eq:rho-pure}
  \rho_i=p_i(s)\ket{\psi_i(s)}\bra{\psi_i(s)},
\:
p_i(s)=\braket{\vc{T}_i(s)\psi}{\vc{T}_i(s)\psi}\equiv N_i^2,
\notag
\\
&
\ket{p_i(s)}=N_i^{-1}\ket{\psi_i(s)},
\:
\ket{\psi_i(s)}\equiv
\ket{\vc{T}_i(s)\psi}.
\end{align}
The phases can now be easily computed
giving the results 
in the following well-known form:
\begin{align}
&
  \label{eq:PhiP-pure}
  \Phi_P = \arg \braket{\psi_1(s)}{\psi_2(s)},
\\
&
\label{eq:PhiI-pure}
\Phi_I=\Phi_g=
\arg \e^{-i \Phi_d(s)}
\frac{\braket{\psi_1(s)}{\psi_2(s)}}{\braket{\psi_1(0)}{\psi_2(0)}},
\end{align}
where $\Phi_d(s)$ is the dynamical phase
given by
\begin{align}
&
\label{eq:Phid-pure}
  \Phi_d(s)=\phi_2(s)-\phi_1(s),
\\
&
\phi_i(s)=-i\int_0^s\bra{p_i(\tau)}\prt{\tau}\ket{p_i(\tau)}\dd
  \tau=
\notag
\\
&
\Im\int_0^s\frac{\bra{\psi_i(\tau)}\prt{\tau}\ket{\psi_i(\tau)}}{\braket{\psi_i(\tau)}{\psi_i(\tau)}}\dd\tau.  
\end{align}

%%%%%%%%%%%%%%
\section{Experiments and Results}
\label{sec:experiment}
%%%%%%%%%%%%%%

In this section,
we describe 
our experiments
performed using
the experimental setup based
on a Mach-Zehnder
two-arm interferometer
and
the results of 
our general theoretical analysis
will be used
to interpret the experimental data.
In these experiments,
the input beam in both arms
is partially polarized
and
effects of
the degree of polarization
will be of our primary concern. 

We consider two
different types of dynamics
represented by
the two experimental configurations:
(a)~the polarization state of 
partially polarized input waves
is modulated by rotation of 
a quarter-wave plate (QWP);
and (b)~the testing (sample) beam
passes through
the cell
filled with a ferroelectric liquid crystal
(FLC) used as an electrically driven light modulator
governed by the orientational Kerr effect~\cite{Kiselev:pre:2:2014}.
In the next subsection we first discuss 
the common elements of our setup 
shared by both of these configurations.  

\begin{figure}[!tbh]
\centering
\resizebox{55mm}{!}{\includegraphics*{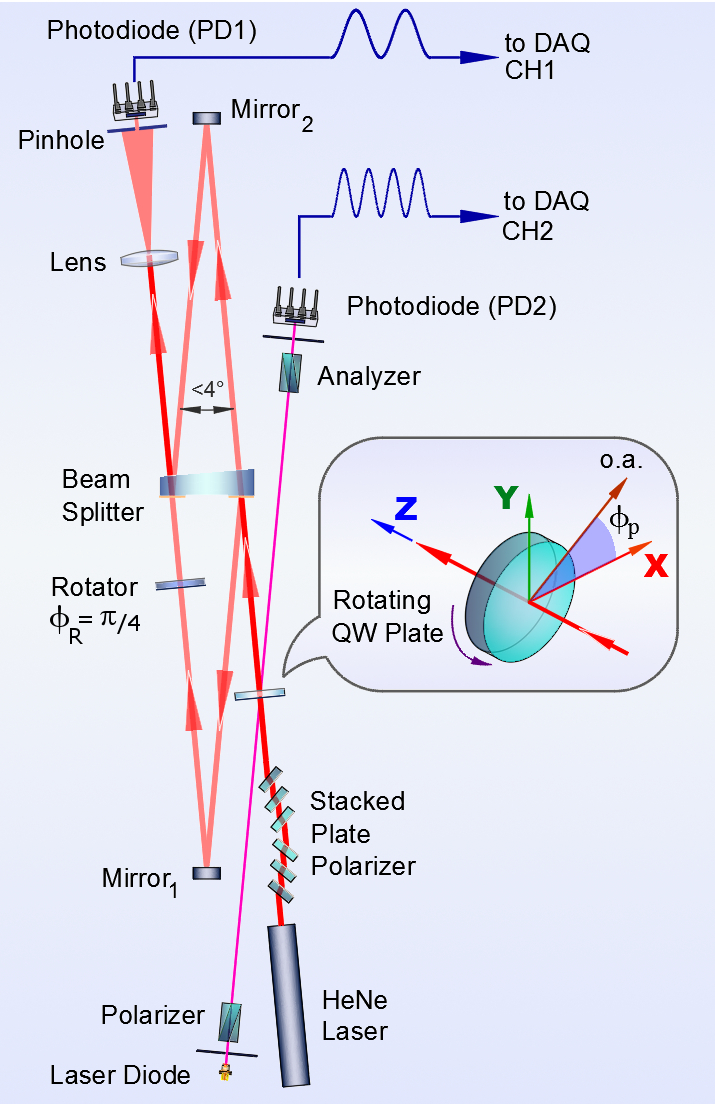}}
\caption{%
Experimental setup 
based on a Mach-Zehnder two-arm interferometer
with the rotating quarter-wave plate
placed before the input beam splitter.
The plate in the sample arm
(rotator) rotates the plane of polarization
of a linear polarized wave
by the angle $\Delta\Phi_R=\pi/4$.
}
\label{fig:mz_qwp}
\end{figure}

%%%%%%%%%%%%%%%%%%
\subsection{Experimental setup} 
\label{subsec:setup}
%%%%%%%%%%%%%%%%%%

In our setup shown in Fig.~\ref{fig:mz_qwp},
a helium-neon laser (the wavelength is $632.8$~nm)
with all polarizing elements removed from the resonator 
was used as a source of unpolarized light.
This light was then converted into
a partially polarized wave 
with the prescribed value of the degree
of polarization, $P_0$,
by means of 
a stacked plate polarizer combined with a series of
dichroic polaroids of variable degree of 
dichroic dye degradation.
The polarized part of this wave was linearly
polarized along the $x$ axis.
Note that each installation of
polarizing elements is followed by 
measurement of the degree of polarization, $P_0$, 
using  a combination of
a rotating analyzer 
and a light detector. 

Referring to Fig.~\ref{fig:mz_qwp},
a beam splitter divides a partially polarized light
into two beams, the reference and the testing (sample) beams,
which, after reflection
at the mirrors $M_1$ and $M_2$, 
are recombined at the semireflecting surface
of  the beam splitter. 
The interfering beams
emerging from the interferometer 
are projected by the lens 
onto a screen with a pinhole
(the diameter was 150~\mum).
After passing the pinhole,
light  is collected by 
a photodiode (PD1)
(the silicon photodiode OTP101 from Texas Instruments)
and the signal is then 
transmitted to a 12-bit data acquisition system (DAQ).
 
The interferometer was adjusted 
to obtain the fringes of equal thickness.
Owing to the elongated geometry of the interferometer,
all the directions of incidence were close to
the normal (deviations from the normal were less than $2^\circ$)
thus making the polarizing effects of Fresnel reflections
negligibly small.  
The period of the interference pattern
was at least $100$ times larger than the pinhole diameter.
Hence errors arising from pinhole induced distortions of the intensity profile
of interference pattern were below $10^{-3}$.
General accuracy
of our measurements of the light intensity
was affected 
by errors resulting from
mechanical instability
and noises of light source,
photodiodes and DAQ. 
This accuracy is estimated to
be below $3.0$\%.

\begin{figure}[!tbh]
\centering
\resizebox{25mm}{!}{\includegraphics*{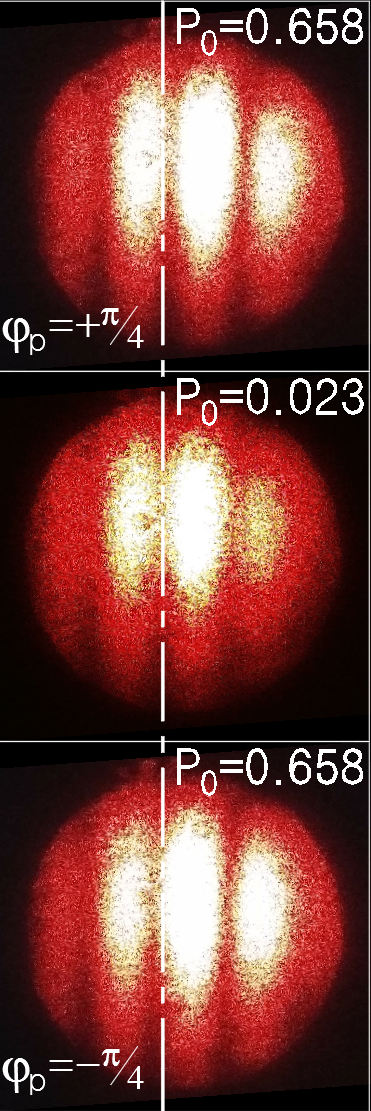}}
\caption{%
The interference fringes 
  observed in the experimental setup 
shown in Fig.~\ref{fig:mz_qwp}.
For nearly-unpolarized light with $P_0=0.023$,
the pattern is independent of $\phi_p$.
For partially polarized light
with $P_0=0.658$, the fringes
at $\phi_p=\pm\pi/4$ are shifted.
}
\label{fig:pattern_qwp}
\end{figure}

%%%%%%%%%%%%%%%%%%%%
\subsection{Interferometer with rotating quarter-wave plate}
\label{subsec:qwp}
%%%%%%%%%%%%%%%%%

%%%%%%%%%%%%%%%
\subsubsection{Experimental procedure}
\label{subsubsec:exper-proc-qwp}
%%%%%%%%%%%%%%%%%%%%%%%

In this setup,
the quarter-wave plate (QWP)
is placed before the beamsplitter
at the input of the interferometer.
This plate is rotated about its normal
(the frequency was fixed at about $5$~Hz)
and orientation
of its in-plane optical axis
specified by 
the azimuthal angle, $\phi_p$, 
continuously changes
leading to
modulation of  the polarization state of 
the partially polarized wave.

In order to get the results
that do not rely on the assumption
of uniformly rotating QWP,
the
azimuthal angle $\phi_p$
was measured
using a crossed polarizers probing scheme
[see Fig.~\ref{fig:mz_qwp}]. 
In this scheme,
the nearly-normal incident probing wave
passes through the QWP placed
between the crossed polarizer and the analyzer
and the intensity of the transmitted beam is registered by
the photodiode (PD2).
Since this intensity
is known to be proportional to $\sin^2(2\phi_p)$,
for our 12-bit DAQ system, 
the azimuthal angle of the QWP optical axis 
was measured with excessively high accuracy limited by the noise. 
 
% (Note, in order to
% avoid collisions, the probing beam propagated with a slight angle to
% the normal of the QWP and this only affects on the modulation depth
% $A_{p}^{2}$ ). 

As is illustrated in Fig.~\ref{fig:mz_qwp},
by contrast to the reference beam,
the testing beam passes through 
the optically active quartz rotator  
that rotates the plane of polarization
by $45$\dega. 
The beams 
are recombined at the beamsplitter
to produce the interference pattern. 

\begin{figure}[!tbh]
\centering
\resizebox{55mm}{!}{\includegraphics*{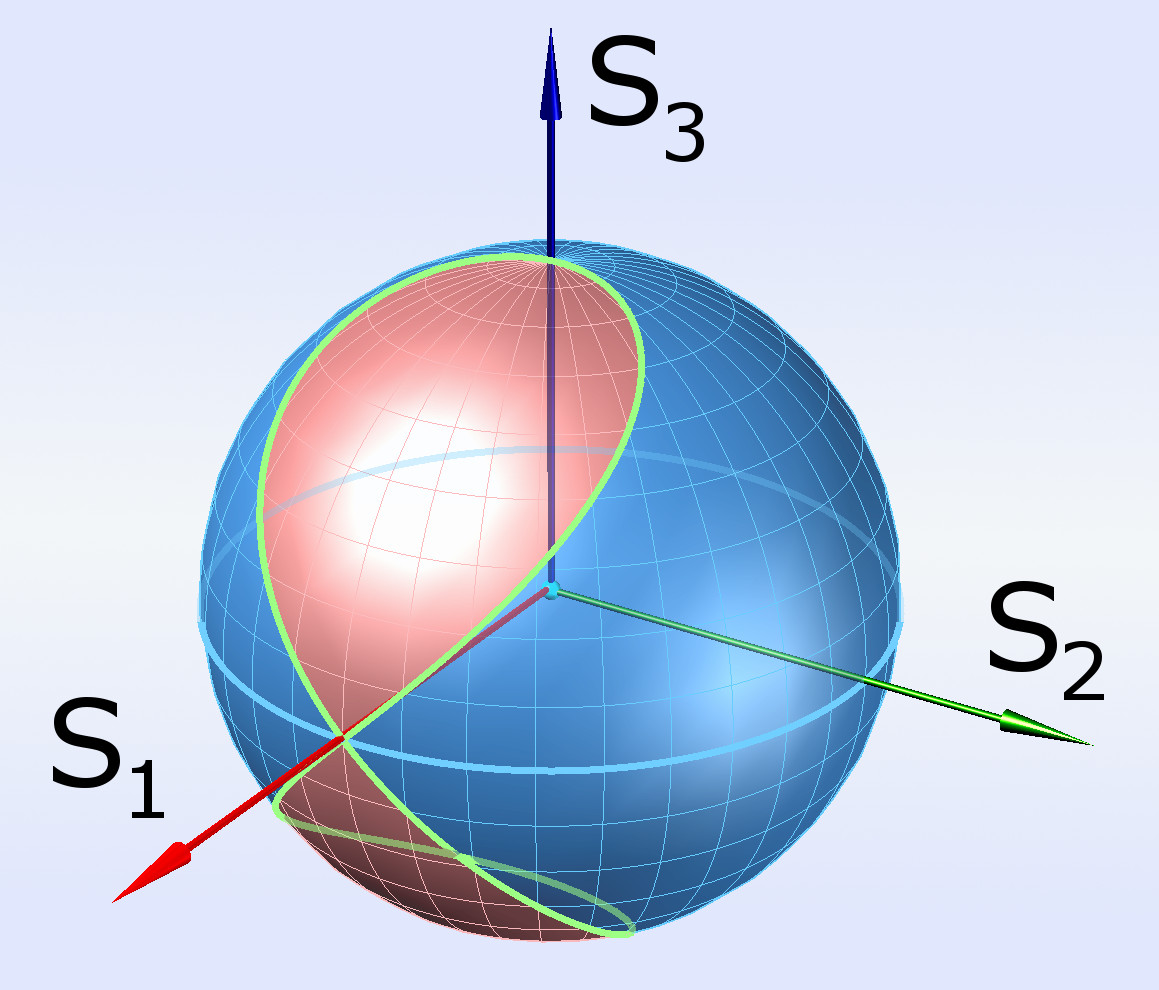}}
\caption{%
Trajectory on the Poincar\'e sphere
shows the Stokes parameters of 
light beam passed through
the quarter-wave plate (QWP) 
as a function of the QWP azimuthal angle. 
}
\label{fig:trajectory_qwp}
\end{figure}

An example of this pattern projected onto the screen 
is shown in Fig.~\ref{fig:pattern_qwp}. 
In general, the contrast and location of the fringes
both depend on the QWP azimuthal angle 
and the degree of polarization, $P_0$.

In order to
register small shifts of the interference pattern
for nearly-unpolarized waves,
we need to maximize sensitivity
of the measurements.
For this purpose, 
the
pinhole is placed 
at the center of the interval
separating adjacent
maxima and minima of the pattern. 
After all preparations, the QW plate was rotated and signals 
from the photodiodes, 
PD1 and PD2,
were registered and digitized in parallel 
by the DAQ (CH1 and CH2).

% The maximum shift was observed
% for a degree of polarization close to unity and for $\phi_p =
% \frac{n\pi}{4}$ (here $n$ is an integer) and was $\pm1/4$ of the
% fringe period from the zero position (where $\phi_p=0$). 

% To this purpose, the QWP stopped at zero azimuth, then the
% screen with pinhole and the PD2 photodiode as a solid block manually
% moved  with the screw feed. The correctness of the positioning was
% controlled by the CH2 signal which level have had to be half of the
% modulation depth. 

\begin{figure}[!tbh]
\centering
\resizebox{85mm}{!}{\includegraphics*{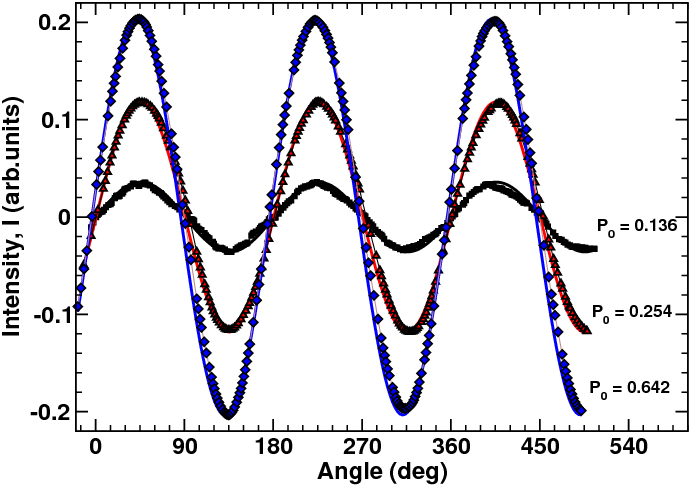}}
\caption{%
Intensity measured as a function
of the QWP azimuthal angle at various
values of the degree of polarization.
Solid lines represent theoretical curves
computed using
Eq.~\eqref{eq:FP_qwp} at $\Delta\Phi_R=\pi/4$. 
}
\label{fig:int_psi}
\end{figure}

%%%%%%%%%%%%%%%
\subsubsection{Results}
\label{subsubsec:results-qwp}
%%%%%%%%%%%%%%%%%%%%%%%

An important point to start from is that,
for these experiments,
the azimuthal angle of the QWP 
optical axis (the QWP angle) 
plays the role of the governing parameter: $\phi_p=s$.
In this subsection,
we  present theoretical analysis of the experiments
along with the experimentally measured data.

By assuming that transmission of light through a birefringent plate
is governed by the unitary matrix of the general form:
\begin{align}
&
  \label{eq:Tpl}
  \vc{T}_p(\phi_p,\eta_p)=\vc{C}(\pi/4,\phi_p)\e^{i\eta_p\bs{\sigma}_3}\hcnj{\vc{C}}(\pi/4,\phi_p)=
\notag
\\
&
\cos(\eta_p)\bs{\sigma}_0+i\sin(\eta_p)\sca{\uvc{m}_p}{\bs{\sigma}},
\end{align}
where $2\eta_p=\Delta\Phi_p$ is the phase retardation;
$\phi_p$ is the azimuthal angle of the in-plane principal axes
and  $\uvc{m}_p=(\cos(2\phi_p),\sin(2\phi_p),0)$,
we can describe how the quarter-wave plate (QWP) 
with $\Delta\Phi_p=\pi/2$
modulates the initial
polarization state~\eqref{eq:rho_0-circ}
with $\uvc{s}_0=\uvc{x}$ 
and obtain
the density matrix
of light passed through the plate
\begin{align}
&
  \label{eq:rho_Tpl}
  2\vc{T}_p\rho_0\hcnj{\vc{T}_p}\equiv 2\rho_p=\bs{\sigma}_0+
P_0 \sca{\uvc{s}_p}{\bs{\sigma}},
\\
&
  \label{eq:mp_qw}
  \uvc{s}_p=\cos(2\phi_p)\uvc{m}_p+\sin(2\phi_p)\uvc{z}=
\notag
\\
&
(\sin(2\theta_p)\cos(2\phi_p),\sin(2\theta_p)\sin(2\phi_p),\cos(2\theta_p)),
\end{align}
where $\theta_p=\pi/4-\phi_p$
and we have used the algebraic result~\eqref{eq:rho_eta}
presented in Appendix~\ref{sec:appendix}.

The effect of QWP rotation can be visualized as 
the trajectory of the normalized Stokes vector $\uvc{s}_p(s)$
given by Eq.~\eqref{eq:mp_qw} 
with $\theta_p=\pi/4-s$ and $\phi_p=s$
on the Poincar\'e sphere.
As is shown in Fig.~\ref{fig:trajectory_qwp},
this trajectory is figure eight shaped
and passes through the poles
representing the states of circular polarization.

The density matrix $\rho_p$
with the Stokes unit vector
given by Eq.~\eqref{eq:mp_qw}
characterizes the reference beam,
$\rho_p=\rho_1$,
and can be recast into the following form:
\begin{align}
&
  \label{eq:rhop}
  2\rho_p(s)\equiv2\rho_1(s)=\vc{D}_p(s)(\bs{\sigma}_0+P_0\bs{\sigma}_3)\hcnj{\vc{D}}_p(s),
\\
&
\label{eq:Dp}
 \vc{D}_p(s)\equiv\vc{D}_1(s)=\vc{D}(\phi_p,\theta_p,\gamma_p)=\vc{D}(s,\pi/4-s,\mp
  s),
\\
&
\label{eq:D_Wigner}
\vc{D}(\alpha,\beta,\gamma)=\e^{-i\alpha\bs{\sigma}_3}\e^{-i\beta\bs{\sigma}_2}\e^{-i\gamma\bs{\sigma}_3}.
\end{align}
The components of 
the eigenpolarization vectors of the reference beam
$\ket{p_{\mu}^{(1)}}$
are given by the columns of the matrix $\vc{D}_p(s)=\vc{D}_1(s)$.
In order to make these vectors well-defined at the poles,
we have introduced the phase factors for the eigenstates
with 
$\gamma_p=-\phi_p$ ($\gamma_p=\phi_p$)
for the upper (lower) half of the Poincar'e sphere.

The testing beam is additionally transmitted through
the rotator described by the transmission matrix:
\begin{align}
&
  \label{eq:U_r}
\vc{T}_R=\e^{i\Phi_R}\vc{U}_R,
\:
  \vc{U}_R=\e^{i\Delta\Phi_R\bs{\sigma}_3}=
\notag
\\
&
\cos\Delta\Phi_R\bs{\sigma}_0+i\sin\Delta\Phi_R\bs{\sigma}_3.
\end{align}
The rotator results in rotation of polarization ellipse axes
by the angle $\Delta\Phi_R$
thus producing the difference
between the reference beam
with the density matrix $\rho_1(s)$
and the testing lightwave
which is characterized by
the density matrix  
\begin{align}
  \label{eq:UrDp}
  \rho_2(\phi_p,\theta_p)=\vc{U}_R\,\rho_1\hcnj{\vc{U}}_R=
\rho_1(\phi_p-\Delta\Phi_R,\theta_p)
\end{align}
 and the matrix of
eigenpolarization vectors
\begin{align}
\label{eq:D2_qwp}
\vc{D}_2=\vc{U}_R\vc{D}_1=\vc{D}(\phi_p-\Delta\Phi_R,\theta_p,\gamma_p).  
\end{align}
The Pancharatnam function~\eqref{eq:PhiP}
can now be readily computed as follows
\begin{align}
&
  \label{eq:FP_qwp}
  \e^{-i\Phi_R}F_P=\Tr[\vc{U}_R\rho_1]=
\notag
\\
&
\cos\Delta\Phi_R+iP_0\sin\Delta\Phi_R\,\cos(2\theta_p),
\end{align}
where $\cos(2\theta_p)=\sin (2s)$.

\begin{figure}[!tbh]
\centering
\resizebox{85mm}{!}{\includegraphics*{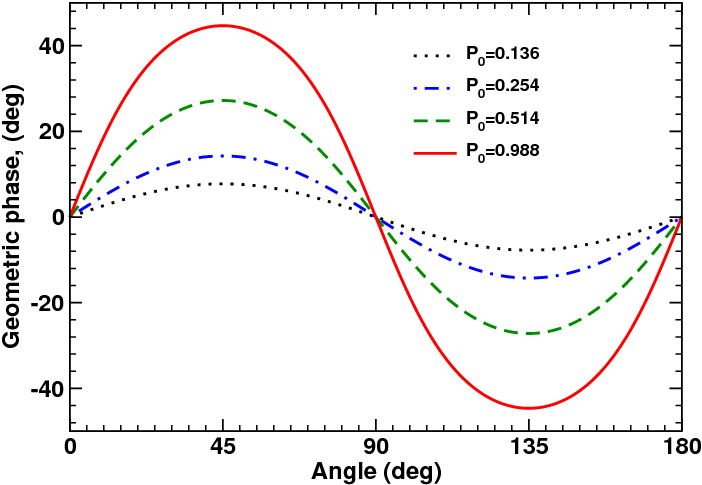}}
\caption{%
Dependence of the relative geometric
phase, $\Phi_I=\Phi_U^{(12)}$,
on the QWP azimuthal angle at different
values of the degree of polarization. 
}
\label{fig:phase_psi}
\end{figure}

We can now 
insert this formula
into 
Eqs.~\eqref{eq:I-tot}--~\eqref{eq:PhiP}
and calculate the angular dependent part of 
the output wavefield intensity.
This is the intensity
which is measured experimentally
as described in
the previous section
and is presented in Fig.~\ref{fig:int_psi}. 
Clearly, the theoretical curves shown
in Fig.~\ref{fig:int_psi} are in excellent agreement with
the experimental data.

\begin{figure}[!tbh]
\centering
\resizebox{85mm}{!}{\includegraphics*{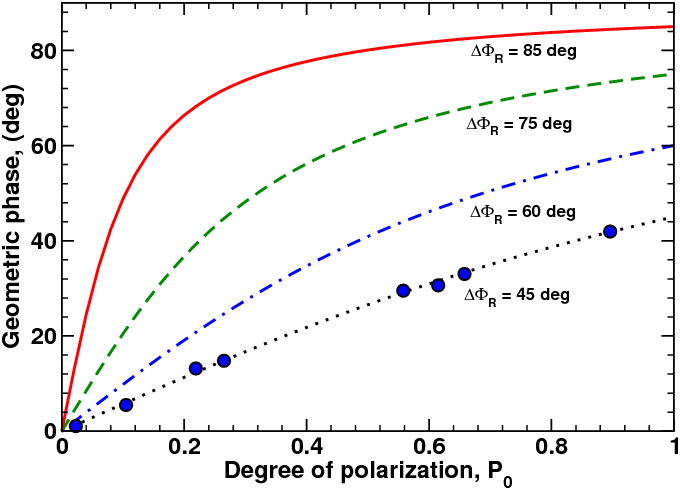}}
\caption{%
Dependence of the relative geometric
phase, $\Phi_I=\Phi_U^{(12)}$, at $s=\pi/4$
on the degree of polarization
at various values of the rotator angle.
Experimental points are marked by solid circles.  
}
\label{fig:phase_P}
\end{figure}

For unitary evolution,
where
$\vc{A}_i=\hcnj{\vc{T}}_i\vc{T}_i=\vc{I}$
and $\tilde{\rho}_i=\rho_0$,
the Pancharatnam function~\eqref{eq:FP_qwp} 
can also be expressed in terms of the eigenstates
\begin{align}
  \label{eq:F_Pi_qwp}
  2F_P=\e^{i\Phi_R}\sum_{\mu=\pm 1}(1+\mu P_0)\braket{p_\mu^{(1)}(s)}{p_\mu^{(2)}(s)},
\end{align}
where
\begin{align}
&
  \label{eq:pmpm-qwp}
  \braket{p_\mu^{(1)}(s)}{p_\mu^{(2)}(s)}
=\frac{1}{2}
\Tr[(\bs{\sigma}_0+\mu\bs{\sigma}_3)\hcnj{\vc{D}}_1\vc{U}_R\vc{D}_1]
=
\notag
\\
&
\cos\Delta\Phi_R+i\mu\sin\Delta\Phi_R \sin(2s),
\end{align}
and the interferometric phase is given by Eq.~\eqref{eq:FI_uni}.
For the dynamical phases, we have
\begin{align}
  \label{eq:phi_d_uni}
  \Phi_\mu^{(d)}=\phi_{\mu}^{(2)}-\phi_{\mu}^{(1)},
\end{align}
where
\begin{align}
  \label{eq:phi_d_qwp}
\phi_{\mu}^{(i)}=
-  \frac{i}{2}
\Tr\Bigl[
(\bs{\sigma}_0+\mu\bs{\sigma}_3)
\int_{0}^s
\hcnj{\vc{D}}_i\prt{\tau}\vc{D}_i\dd\tau
\Bigr].
\end{align}
Equation~\eqref{eq:D2_qwp}
shows that the eigenpolarization vectors
of the waves are related through the unitary
$\vc{U}_R$.
Since this unitary is independent
of the evolution parameter $s$,
the operators
$\hcnj{\vc{D}}_1\prt{\tau}\vc{D}_1$
and $\hcnj{\vc{D}}_2\prt{\tau}\vc{D}_2$
[formulas for these operators
are given in Eq.~\eqref{eq:H_D} of Appendix~\ref{sec:appendix}]
are identical and
the difference between
the dynamical phases 
$\phi_{\mu}^{(1)}$ and $\phi_{\mu}^{(2)}$ 
vanishes.
So, the resulting expression for
the interferometric phase is
\begin{align}
  \label{eq:Phig-qwp}
  \Phi_I=\arg[\cos\Delta\Phi_R+iP_0\sin\Delta\Phi_R\,\cos(2s)]
=\Phi_P-\Phi_R.
\end{align}

Now we discuss the relative Uhlmann phase
given by Eq.~\eqref{eq:PhiU_rel}.
For this purpose,
we shall apply the results
of Appendix~\ref{sec:appendix}
to our case, where $\vc{T}_1=\vc{T}_p$
and $\vc{T}_2=\vc{U}_R\vc{T}_1$.
After substituting $\eta=\pi/4$
and $\psi=2 s$ into
the relation~\eqref{eq:H_0},
we have
\begin{align}
&
  \label{eq:h_p}
  \hcnj{\vc{T}}_1\prt{s}\vc{T}_1=\hcnj{\vc{T}}_2\prt{s}\vc{T}_2=
i\sca{\vc{h}_p}{\bs{\sigma}},
\notag
\\
&
\vc{h}_p=(-\sin(2s),\cos(2s),1).
\end{align}
At $\uvc{s}_0=\uvc{x}$ and $\vc{h}=\vc{h}_p$,
formulas~\eqref{eq:HU}--~\eqref{eq:hU} for the Uhlmann connection
take the form:
\begin{align}
&
  \label{eq:H_U_qwp}
  \vc{H}_U^{(1)}=\vc{H}_U^{(2)}=\sca{\vc{h}_U}{\bs{\sigma}},
\\
&
\label{eq:h_U_qwp}
\vc{h}_U=(1-P_0^2)^{1/2}(\sin(2s)\uvc{x}-\uvc{z})-\cos(2s)\uvc{y}.
\end{align}
From these results it is clear that
the Uhlmann holonomies,
$\vc{V}_U^{(1)}$ and $\vc{V}_U^{(2)}$,
are equal and 
$\vc{V}_U^{(12)}=\vc{I}$.
Formulas~\eqref{eq:PhiU_rel}--~\eqref{eq:V_rel}
then give the relative Uhlmann phase
in the simple form:
\begin{align}
  \label{eq:Phi_U_qwp}
  \Phi_U^{(12)}(s)=\Phi_P(s)-\Phi_P(0).
\end{align}
Clearly, it means that 
the Uhlmann phase~\eqref{eq:Phi_U_qwp}
is equal to the interferometric phase~\eqref{eq:Phig-qwp}:
$\Phi_U^{(12)}(s)=\Phi_I(s)$.

This is, however, no longer the case
for the relative Uhlmann phase,
$\tilde{\Phi}_U^{(12)}(s)$
given by Eqs.~\eqref{eq:tPhiU_rel}--\eqref{eq:tV_rel}.
For this phase, the operator
$\vc{V}_{12}(0)=\hcnj{\vc{T}}_2(0)\vc{T}_1(0)=\hcnj{\vc{T}}_1(0)\hcnj{\vc{U}}_R\vc{T}_1(0)$
that enter Eq.~\eqref{eq:tV_rel}
differs from the identity matrix 
preventing the contributions from the Uhlmann connections
of the beams from being canceled out.

In Fig.~\ref{fig:phase_psi}, we present
dependence of the relative geometric phase
on the QWP azimuthal angle
computed from Eq.~\eqref{eq:Phig-qwp} at various values of the
degree of polarization.
As it can be seen from Fig.~\ref{fig:phase_psi},
this dependence is perfectly harmonic
and the geometric phase
is maximal at $\phi_p\equiv s=\pi/4$.

Figure~\ref{fig:phase_P} shows that
the maximum value of the geometric
phase is an increasing function of 
the degree of polarization.
Referring to Fig.~\ref{fig:phase_P},
this dependence becomes more pronounced
as the rotator angle $\Delta\Phi_R$
approaches $\pi/2$.
Though such rotator will suffer from 
nearly-zero contrast
of the fringes.

\begin{figure}[!tbh]
\centering
\resizebox{55mm}{!}{\includegraphics*{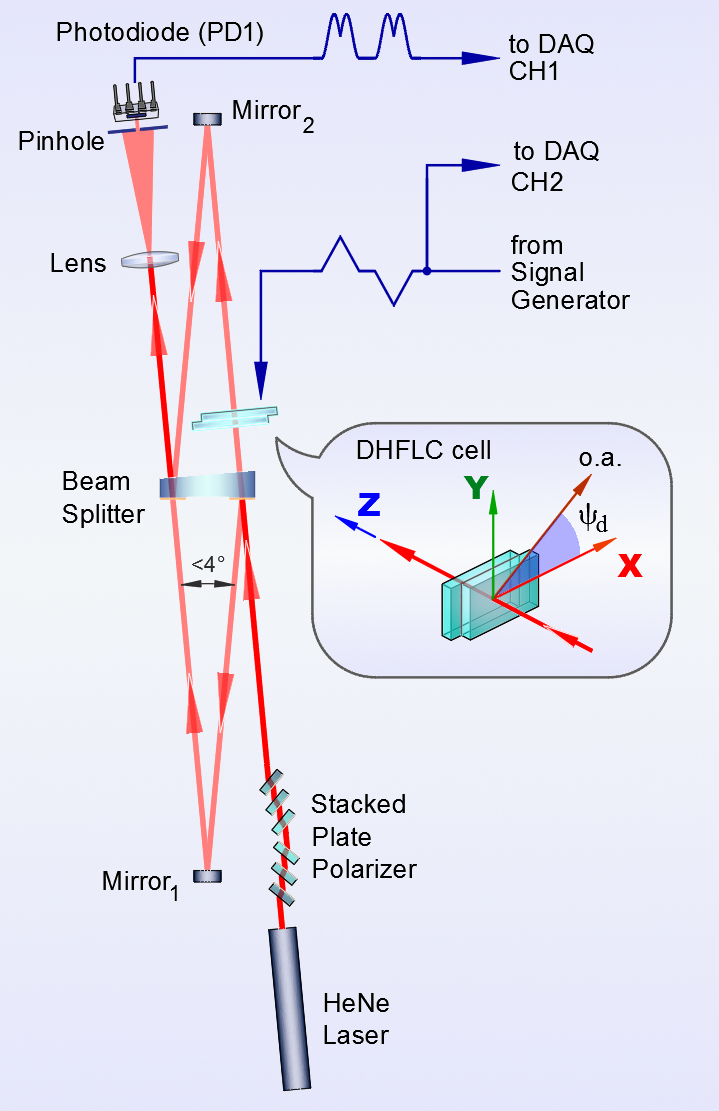}}
\caption{%
Experimental setup 
based on a Mach-Zehnder two-arm interferometer
with the DHFLC cell (filled with
ferroelectric liquid crystal)
placed into the testing arm.
}
\label{fig:mz_dhf}
\end{figure}

%%%%%%%%%%%%%%%%%%%%
\subsection{Interferometer with DHFLC cell}
\label{subsec:dhflc}
%%%%%%%%%%%%%%%%%

In our previous studies~\cite{Kiselev:pre:2015,Kiselev:pre:2017},
we have employed 
the experimental setup
based on a Mach-Zehnder
interferometer to study electro-optic
response of 
planar aligned 
deformed-helix ferroelectric liquid crystals
(DHFLC) with subwavelength helix pitch. 

In such ferroelectric liquid crystals (FLCs),
the equilibrium orientational structure
forms a helical twisting pattern
where FLC molecules align on average along
a local unit director
$
\uvc{d}=
\cos\theta\,\uvc{h}+
\sin\theta\,\uvc{c},
%\label{eq:director}
$
where $\theta$ is the {smectic tilt angle}; 
$\uvc{h}=\uvc{x}$ is the twisting axis normal to the smectic layers and
$\uvc{c}\perp\uvc{h}$ is the $c$-director.
The FLC director lies on the smectic cone 
and rotates
in a helical fashion about a uniform twisting axis
$\uvc{h}$ forming the FLC helix. 
The smectic layers are normal to the substrates
and the electric field $\vc{E}=E\,\uvc{z}$ 
is applied across the DHFLC cell.

According to
Refs.~\cite{Kiselev:pre:2011,Kiselev:pre:2:2014,Kiselev:pre:2015,Kiselev:jmolliq:2018},
optical properties of such cells
can be described by the effective dielectric tensor
of a homogenized DHFLC helical structure.
The zero-field 
($\mathbf{E}=0$)
dielectric tensor is
uniaxially anisotropic with 
the optical axis directed along the
twisting axis $\hat{\mathbf{h}}=\hat{\mathbf{x}}$.
The zero-field effective refractive indices of
extraordinary (ordinary) waves, $n_h$ ($n_p$),
generally depend
on the smectic tilt angle
$\theta$ and
the optical dielectric
constants
characterizing the FLC material
(see, e.g., equation (56) in
Ref.~\cite{Kiselev:pre:2:2014}
giving the expressions for $\epsilon_h=n_h^2$
and $\epsilon_p=n_p^2$).

The electric-field-induced
anisotropy is generally biaxial
so that the dielectric tensor
is characterized by the three generally different principal values:
$\epsilon_{\pm}=n_{\pm}^2$ and $\epsilon_z=n_z^2$
(see, e.g., equations (60)--(63) in
Ref.~\cite{Kiselev:pre:2:2014}).
The in-plane principal optical axes,
$\hat{\mathbf{d}}_{+}=\cos\psi_d\,\hat{\mathbf{x}}+
\sin\psi_d\,\hat{\mathbf{y}}$
and
$\hat{\mathbf{d}}_{-}=\hat{\mathbf{z}}\times  \hat{\mathbf{d}}_{+}$,
are rotated about the
vector of electric field, $\mathbf{E}\parallel \hat{\mathbf{z}}$,
by the azimuthal angle $\psi_{\mathrm d}$
(see, e.g., equation (64) in
Ref.~\cite{Kiselev:pre:2:2014}).
This is the  so-called orientational Kerr effect
which is caused by
the electrically induced distortions of the helical
structure
and is governed by the effective dielectric tensor
of a nanostructured chiral smectic liquid crystal
defined through averaging over 
the FLC orientational 
structure~\cite{Kiselev:pre:2013,Kiselev:ol:2014,Kiselev:pre:2011,Kiselev:pre:2:2014}.

%%%%%%%%%%%%%%%
\subsubsection{Experimental procedure}
\label{subsubsec:exper-proc-dhflc}
%%%%%%%%%%%%%%%%%%%%%%%

Our experimental setup 
with the DHFLC cell placed in the testing arm
is depicted in Fig.~\ref{fig:mz_dhf}.
Orientation of the cell 
was  adjusted so as to have
the zero-field optical axis
parallel to the 
$x$ axis
(the polarized part of incident lightwave is also linearly
polarized along the $x$ axis).
The mirrors were fine tuned
so as to position the pinhole at the center of a dark fringe
in the field-free interference pattern. 

Measurements were performed
for triangular wave-form of driving voltage
with the frequency $f=100$~Hz
(alternating
triangular pulses with a duty cycle 4/6 ($-\wedge-\vee$)).
For this purpose, the signal registered by the photodiode (PD1)
was recorded and digitized in parallel
with the signal from the generator
using  the same DAQ (by CH1 and CH2). 

\begin{figure}[!tbh]
\centering
\resizebox{85mm}{!}{\includegraphics*{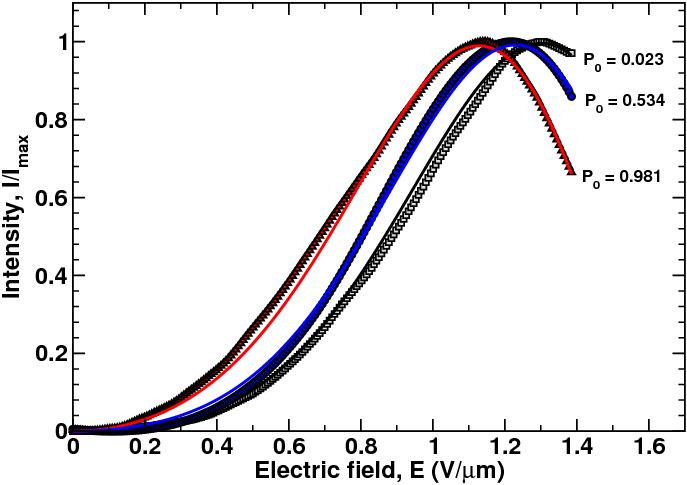}}
\caption{%
Normalized intensity of output wavefield 
experimentally measured 
as a function of the electric field 
at different values of the degree of
polarization, $P_0$,
for the
DHFLC cell of thickness 
$D\approx 52$~\mum\ 
filled with the FLC mixture FLC-587F7.
Solid lines represent the theoretical
curves. 
The parameters of the mixture are: $n_{\perp}\approx 1.47$ is the ordinary refractive
index, $n_{\parallel}\approx 1.69$ is the extraordinary refractive
index, 
$\theta\approx 33.5$\dega\  
is the smectic tilt angle,
and $r_2\approx 1.05$ is the biaxiality ratio.
}
\label{fig:expt_dhf}
\end{figure}

Electric field dependence of 
the  light intensity measured 
at different values of the degree of polarization
are presented in Fig.~\ref{fig:expt_dhf}.
In these experiments we have used the FLC mixture FLC-587F7
(from P.N. Lebedev Physical Institute of Russian Academy of
Sciences) as a FLC material for the DHFLC cell
(a similar mixture was detailed in Ref.~\cite{Kiselev:pre:2015})
% The FLC-587F7 layer is sandwiched between two glass substrates
% covered by indium tin oxide (ITO) and aligning films with 
% the thickness
% $20$~nm and the gap is fixed by spacers at 
% $D\approx 52$~$\mu$m. 

%%%%%%%%%%%%%%%
\subsubsection{Results}
\label{subsubsec:results-dhflc}
%%%%%%%%%%%%%%%%

In the setup shown in Fig.~\ref{fig:mz_dhf},
the testing beam propagates through
the DHFLC cell which can be regarded
as an electrically driven birefringent plate
with the unitary transmission matrix
$\mvc{T}$ given by 
\begin{align}
&
  \label{eq:T-FLC}
\vc{T}_1=\mvc{I},\:
  \vc{T}_2=\mvc{T}(\tau)=
\vc{C}_d
\begin{pmatrix}
    \e^{i\Phi_{+}} & 0\\
    0 &\e^{i\Phi_{-}} 
  \end{pmatrix}.
\hcnj{\vc{C}}_d=
\notag
\\
&
\e^{i\Phi}\vc{T}_p(\psi_d,\Delta\Phi)=\e^{i\Phi}\e^{i\Delta\Phi(\uvc{s}_d\cdot\bs{\sigma})},
\end{align}
where
$\vc{C}_d=\vc{C}(\pi/4,\psi_d)$
and
$\Phi_{\pm}=n_{\pm} \tau$;
$\Phi=(\Phi_{+}+\Phi_{-})/2$
is the averaged phase shift;
$2\Delta\Phi=(n_{+}-n_{-})\tau$ 
is the difference in 
optical path of the ordinary and extraordinary waves known
as the phase retardation;
$\psi_d$ is the azimuthal angle of the in-plane
optical axis and
$\uvc{s}_d=(\cos(2\psi_d),\sin(2\psi_d),0)$.
In contrast to the previous section,
the governing parameter now is 
the thickness parameter 
of the DHFLC cell, $0\le \tau\le h\equiv k_{\ind{vac}}D$,
where $k_{\ind{vac}}=\omega/c$ is the free-space wave number
and $D$ is the cell thickness.

Thus we have the simple case
of unitary evolution with 
$\vc{T}_1=\vc{T}(0)=\vc{I}$
and
the Pancharatnam function is given by
\begin{align}
&
  \label{eq:F_P-flc}
F_P(h)=
\Tr[\rho_0\vc{T}(h)]=
\e^{i\Phi}(\cos(\Delta\Phi)+i P_0
  \sin(\Delta\Phi)
\notag
\\
&
\times\sca{\uvc{s}_d}{\uvc{s}_0})
=\frac{1}{2}\sum_{\mu=\pm 1}(1+\mu P_0)\braket{p_\mu(0)}{p_\mu(h)},
\end{align}
where $\ket{p_{\mu}(h)}=\vc{T}(h)\ket{p_{\mu}(0)}$
is the eigenpolarization vector
of the density matrix $\rho_{2}(h)=\vc{T}(h)\rho_0\hcnj{\vc{T}}(h)$.

We can now combine the Pancharatnam function~\eqref{eq:F_P-flc}
with the known analytical results for $\Phi_{\pm}$
proportional to the effective refractive indices $n_{\pm}$
and the azimuthal angle $\psi_d$
describing orientation of the in-plane optical axis
(see, e.g., Refs.~\cite{Kiselev:pre:2017,Kiselev:pre:2015,Kiselev:pre:2:2014})
to evaluate the intensity of the output field~\eqref{eq:I-tot}
and to fit the experimental data
on electric field dependence of the
light intensity. 
Figure~\ref{fig:expt_dhf}
presents the experimental results
measured by the photodiode 
at different values of the degree of
polarization, $P_0$, for the
DHFLC cell of thickness 
$D\approx 52$~\mum\ 
filled with the FLC mixture FLC-587F7.
 
The parameters of the mixture 
used in the fitting procedure
are: $n_{\perp}=\sqrt{\epsilon_{\perp}}\approx 1.47$ is the ordinary refractive
index, $n_{\parallel}\approx 1.69$ is the extraordinary refractive
index and $\theta\approx 33.5$\dega\  
is the smectic tilt angle.
Then the fitting
gives the values of two ratios
regarded as the fitting parameters:
the ratio  
of the ferroelectric polarization
and the dielectric susceptibility of the Goldstone mode
$P_s/\chi_E\approx 5.43$~V/\mum\ 
and the biaxiality ratio $r_2=\epsilon_2/\epsilon_{\perp}\approx
1.05$, where $\epsilon_2$ is the principal value
of the dielectric tensor along the FLC polarization vector 
In Fig.~\ref{fig:expt_dhf}, 
the theoretical curves are shown as solid lines. 

\begin{figure}[!tbh]
\centering
\resizebox{85mm}{!}{\includegraphics*{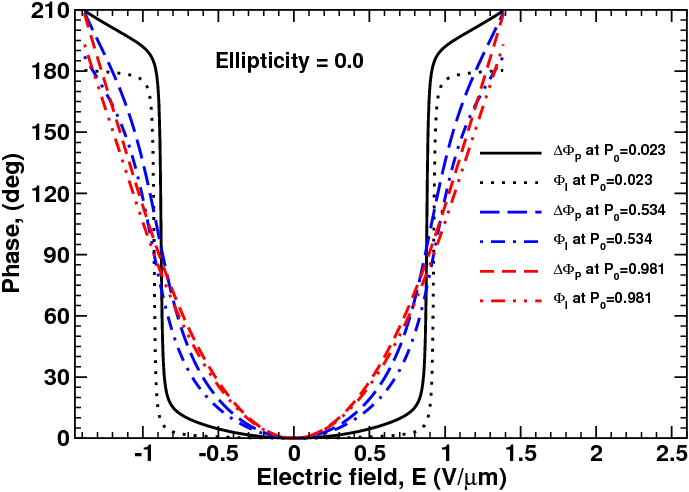}}
\caption{%
Pancharatnam and interferometric phases,
$\Delta\Phi_P=\Phi_P(E)-\Phi_P(0)$
and $\Phi_I$,
as a function of the electric field 
at different values of the degree of
polarization, $P_0$.
}
\label{fig:phi_I_dhf}
\end{figure}

%%%%%%%%%%%%%%%%%%%%%%%
\subsubsection{Geometric phases}
\label{subsubsec:geom-phases-dhflc}
%%%%%%%%%%%%%%%%%%%%%%%

Now we proceed with the geometric phases.
Our task is to compute the phases
as a function of the applied electric field
by using the experimental data combined
with the results of fitting obtained
in Sec.~\ref{subsubsec:results-dhflc}.

\begin{figure}[!tbh]
\centering
\resizebox{85mm}{!}{\includegraphics*{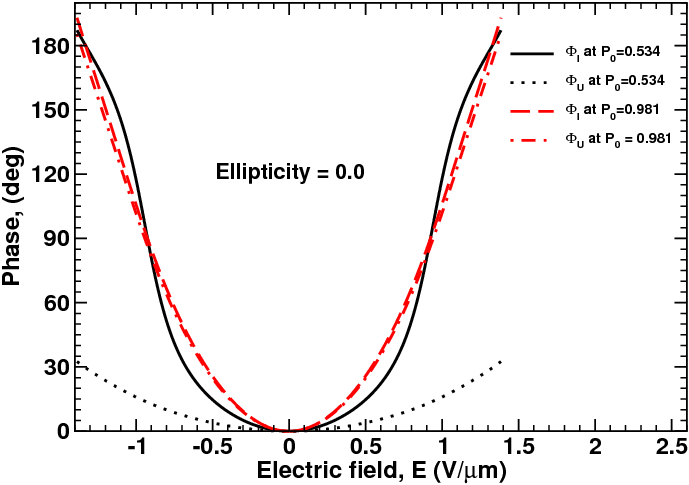}}
\caption{%
Electric field dependence of
Uhlmann and interferometric phases,
$\Phi_U$ and $\Phi_I$,
at different values of the degree of
polarization, $P_0$.
}
\label{fig:phi_U_dhf}
\end{figure}

We begin with calculation
of the interferometric phase.
For this purpose,
we compute the inner products 
\begin{align}
&
  \label{eq:pmpm-flc}
  \braket{p_\mu(0)}{p_\mu(h)}
=\frac{1}{2}
\Tr[(\bs{\sigma}_0+\mu\bs{\sigma}_3)\hcnj{\vc{C}}_0\vc{T}(h)\vc{C}_0]
=
\notag
\\
&
\e^{i\Phi}(\cos\Delta\Phi+i\mu\,\sin\Delta\Phi \sca{\uvc{s}_d}{\uvc{s}_0}),
\end{align}
and the dynamical phases
\begin{align}
&
  \label{eq:phi_dyn-flc}
  \Phi_{\mu}^{(d)}=
-\frac{i}{2}
\int_0^h
\Tr[(\bs{\sigma}_0+\mu\bs{\sigma}_3)\hcnj{\vc{C}}_0\hcnj{\vc{T}}(\tau)\dot{\vc{T}}(\tau)\vc{C}_0]
\dd\tau
\notag
\\
&
=-i\int_0^h\bra{p_{\mu}(\tau)}\prt{\tau}\ket{p_\mu(\tau)}\dd\tau=\Phi+\mu\,\Delta\Phi \sca{\uvc{s}_d}{\uvc{s}_0},
\end{align}
where we have used the identity:
$-i\hcnj{\vc{T}}\dot{\vc{T}}=\vc{H}=\dot\Phi\,\bs{\sigma}_0+\Delta\dot\Phi\,
\sca{\uvc{s}_d}{\bs{\sigma}}$
for the transmission matrix~\eqref{eq:T-FLC},
that enter the interferometric function
\begin{align}
  \label{eq:F_I_flc}
  2F_I=\sum_{\mu=\pm 1}(1+\mu P_0)\braket{p_\mu(0)}{p_\mu(h)}\e^{-i\Phi_{\mu}^{(d)}}.
\end{align}
After substituting 
formulas~\eqref{eq:pmpm-flc}
and~\eqref{eq:phi_dyn-flc}
into Eq.~\eqref{eq:F_I_flc},
we obtain the interferometric phase
in the following form:
\begin{align}
&
  \label{eq:Phi_I_flc}
  \Phi_I=\arg(\Re \tilde{F}_I+i P_0 \Im \tilde{F}_I),
\\
&
\label{eq:tF_I_flc}
\tilde{F}_I=[\cos(\Delta\Phi)+i \sin(\Delta\Phi)\sca{\uvc{s}_d}{\uvc{s}_0}]
\e^{-i\Delta\Phi(\uvc{s}_d\cdot\uvc{s}_0)}. 
\end{align}
The phases~\eqref{eq:phi_dyn-flc}
can also be used to evaluate 
the interferometric holonomy~\eqref{eq:VI_rel}
at $\vc{V}_I^{(2)}=\vc{I}$
and $\ket{r_{\mu}^{(i)}}=\ket{p_\mu(0)}$. 
The result reads
\begin{align}
  \label{eq:V_I_flc}
  \vc{V}_I=\e^{-i\Phi}\e^{-i\Delta\Phi(\uvc{s}_d\cdot\uvc{s}_0)(\uvc{s}_0\cdot\bs{\sigma})}.
\end{align}

\begin{figure}[!tbh]
\centering
\resizebox{85mm}{!}{\includegraphics*{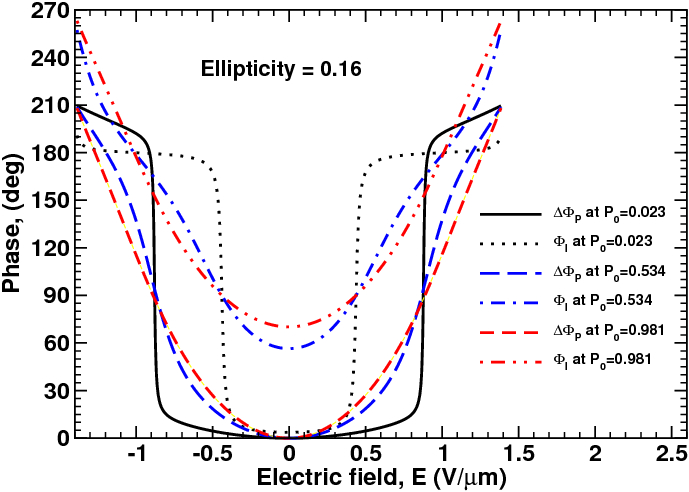}}
\caption{%
Electric field dependence of
the Pancharatnam and interferometric phases,
$\Phi_P$ and $\Phi_I$,
for elliptically polarized waves with
$\epsilon_{\ind{ell}}=\tan(\pi/4-\theta_0)\approx
0.16$
($\theta_0=36\dega$)
at different values of the degree of
polarization, $P_0$.
}
\label{fig:phi_I_016_dhf}
\end{figure}

Now we turn to the Uhlmann phase.
By applying the results of
Appendix~\ref{sec:appendix}
to our case with
$\vc{H}=\dot\Phi\,\bs{\sigma}_0+\Delta\dot\Phi\,
\sca{\uvc{s}_d}{\bs{\sigma}}$
we deduce
the Uhlmann connection
\begin{align}
  &
  \label{eq:H_U_flc}
  \vc{H}_U=-\dot\Phi\bs{\sigma}_0-\Delta\dot\Phi\sca{\vc{q}_U}{\bs{\sigma}},
\\
  \label{eq:q_U_flc}
&
\vc{q}_U=(1-P_0^2)^{1/2}[\uvc{s}_d-\sca{\uvc{s}_d}{\uvc{s}_0}\uvc{s}_0]+
\sca{\uvc{s}_d}{\uvc{s}_0}\uvc{s}_0,
\end{align}
where
$|\vc{q}_U|^2=1-P_0^2(1-\sca{\uvc{s}_d}{\uvc{s}_0}^2)$,
and the corresponding Uhlmann holonomy
\begin{align}
  \label{eq:V_U_flc}
  \vc{V}_U=\e^{-i\Phi}\e^{-i\Delta\Phi(\vc{q}_U\cdot\bs{\sigma})},
\end{align}
so that the Uhlmann phase is given by
\begin{align}
&
  \label{eq:F_U_flc}
\Phi_U  = \arg F_U^{(12)}\equiv\arg F_U
\notag
\\
&
=
\arg
\Tr
[
\sqrt{\rho_0}\,
\e^{i\Delta\Phi(\uvc{s}_d\cdot\bs{\sigma})}
\sqrt{\rho_0}\,
\e^{-i\Delta\Phi(\vc{q}_U\cdot\bs{\sigma})}
].
\end{align}
As is evident from a comparison between
formulas~\eqref{eq:V_I_flc}
and~\eqref{eq:V_U_flc},
the holonomies become identical
in the limiting case of pure states
with $P_0=1$.
It is also clear that, 
by contrast to the interferometric holonomy~\eqref{eq:V_I_flc},
the Uhlmann holonomy~\eqref{eq:V_U_flc} 
depends on the degree of polarization
and does not commute with the density matrix $\rho_0$.

\begin{figure}[!tbh]
\centering
\resizebox{85mm}{!}{\includegraphics*{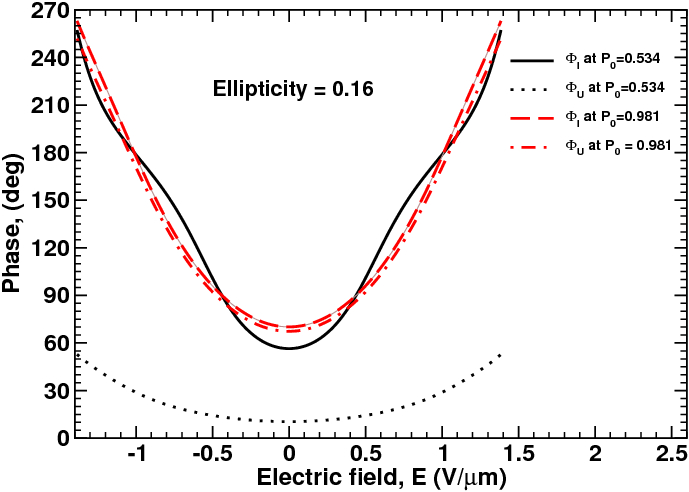}}
\caption{%
Electric field dependence of
the Uhlmann and interferometric phases,
$\Phi_U$ and $\Phi_I$,
for elliptically polarized waves with
$\epsilon_{\ind{ell}}=\tan(\pi/4-\theta_0)\approx
0.16$
($\theta_0=36\dega$)
at different values of the degree of
polarization, $P_0$.
}
\label{fig:phi_U_016_dhf}
\end{figure}

Figure~\ref{fig:phi_I_dhf}
shows how
the Pancharatnam and interferometric
phases 
depend on the electric field
applied across the DHFLC cell.
The curves presented in Fig.~\ref{fig:phi_I_dhf}
are computed from
Eq.~\eqref{eq:F_P-flc}
and Eq.~\eqref{eq:Phi_I_flc}
using the parameters
of the DHFLC cell
that was derived by fitting
the experimental results
[see Fig.~\ref{fig:expt_dhf}].
As can be seen from Fig.~\ref{fig:phi_I_dhf},
the electrically induced parts
of the phases 
are close to each other
and the difference between the curves
decreases with the degree of polarization.

The results for the Uhlmann 
and the interferometric phases
are shown
in Fig.~\ref{fig:phi_U_dhf}.
By contrast to the interferometric phase,
the Uhlmann phase is negligibly
small when the incident
wave is nearly-unpolarized
with $P_0=0.023$.
As it has already been mentioned,
the phases are coincident in the limiting case of 
fully polarized waves with $P_0=1$
and, as is shown
in Fig.~\ref{fig:phi_U_dhf},
there is little difference between 
the curves at $P_0=0.981$.

Ellipticity
of the polarized part of the input
wave,
$\epsilon_{\ind{ell}}=\tan(\pi/4-\theta_0)$,
is determined by the polar angle,
$\theta_0$,
of the Stokes
vector~\eqref{eq:s0-1}.
Figures~\ref{fig:phi_I_016_dhf}
and~\ref{fig:phi_U_016_dhf}
demonstrate what happen to the
phases at non-vanishing
ellipticity.
Referring to
Fig.~\ref{fig:phi_I_016_dhf},
in contrast to the case
of linear polarization,
the curves for
the Pancharatnam and the
interferometric
phases are no longer
close to each other.
From similar curves for
the Uhlmann phase
shown in
Fig.~\ref{fig:phi_U_016_dhf}
it might be concluded
that $\Phi_I$ is
the phase most sensitive
to variations in the ellipticity
as compared to the Pancharatnam and
Uhlmann phases.

In closing this subsection, note that,
in Appendix~\ref{sec:appendix2},
we discuss how to take into account 
the effects of dissipation in birefringent plates
and derive the analytical results
for the Uhlmann and interferometric
phases.

%%%%%%%%%%%%%%
\section{Discussion and conclusions}
\label{sec:conclusion}
%%%%%%%%%%%%%%

In this paper, 
we have studied 
the Uhlmann and interferometric
phases of mixed polarization states
within the framework
of the interferometry of
partially polarized waves.
This interferometry based approach 
assumes that 
the partially polarized beams
independently evolve and, 
depending on their dynamics (evolution),
acquire
both dynamical and geometric phases.
These beams brought together
emerge from the interferometer
producing the interference pattern
where loci of fringes
are determined by 
the Pancharatnam phase
expressed in terms of the Pancharatnam function~\eqref{eq:PhiP}.

Recasting this function into the form
of inner product of the amplitudes 
(purifications) gives the relation~\eqref{eq:PPh_inner}
linking the Uhlmann approach
and the interference pattern.
The relative Uhlmann phase~\eqref{eq:F_U}
defined in terms of the parallel-transported amplitudes~\eqref{eq:w_i}
then appears as a natural consequence
of this relation.

Alternatively, the Pancharatnam function
can be written as a superposition of
inner products of the eigenstates~\eqref{eq:P-phase-1}.
We have found that the structure
of this superposition is complicated
in the presence of dissipation
(nonunitary evolution).
A simpler form 
of this superposition
(see, e.g., Eq.~\eqref{eq:F_P-flc})
was used as a starting point of 
the method giving the interferometric
phase for both types of 
evolution~\cite{Sjoqvist:prl:2000,Tong:prl:2004}.
Following this method, 
we have eliminated 
the dynamical phases of the eigenstates
and derived the interferometric phase~\eqref{eq:PhiI_rel}.
Similar form of our key results for
the relative Uhlmann phase
[see Eqs.~\eqref{eq:PhiU_rel}--~\eqref{eq:V_rel}]
and the interferometric phase
[see Eqs.~\eqref{eq:PhiI_rel}--\eqref{eq:VI_rel}]
clearly indicate that the Uhlmann and interferometric holonomies
determined by different parallel transport conditions
[see Eq.~\eqref{eq:PTU} and Eq.~\eqref{eq:PT_int}]
are responsible for differences between the phases.

In our experimental investigation
into the effects of the degree of polarization
in modulation of partially polarized light
we have employed
the well-known technique
based a Mach-Zehnder two-arm interferometer.
Two different dynamical regimes of light modulation
are studied:
(a)~modulation of the input wave
by the rotating quarter-wave plate
(see Fig.~\ref{fig:mz_qwp});
and
(b)~modulation of the testing beam
by the deformed-helix ferroelectric liquid crystal cell
that can be viewed as a birefringent plate
with electrically controlled anisotropy
(see Fig.~\ref{fig:expt_dhf}).

In the setup using the rotating QWP,
modulation dynamics is governed by
the azimuthal angle of the QWP transmission axis
$\phi_p$ which is regarded as the governing parameter
and the normalized Stokes vector describing 
the polarized part of the input wave
(see Eq.~\eqref{eq:mp_qw})
moves along the figure eight shaped
trajectory on the Poincar\'e sphere
as the QWP angle varies
(see Fig.~\ref{fig:trajectory_qwp}).
The testing beam additionally passes
through the rotator made of quartz
that rotates the plane of polarization 
by the angle $\Delta\Phi_R=\pi/4$.
The interference pattern is
found to be determined by
the Pancharatnam function~\eqref{eq:F_Pi_qwp}
which is used to perform computations
giving the output in excellent agreement with
the experimental data.
Our analysis led to the conclusion that
the interferometric and the relative Uhlmann phases
are equal and
the geometric phase is the phase  
extracted from the results of our measurements
using the Pancharatnam function~\eqref{eq:F_Pi_qwp}.
Figures~\ref{fig:phase_psi}
and~\ref{fig:phase_P}
show how the geometric phase
depends on the QWP angle and
the degree of polarization, respectively.

In our experiments
where the partially polarized testing light is modulated
by the DHFLC cell
light modulation takes place due to 
the orientational Kerr effect in 
ferroelectric liquid 
crystals.
The DHFLC cell regarded
as an optical device
represents
a birefringent plate
with electrically controlled anisotropy
and the physics of the orientational Kerr
effect defines how the phase shifts
($\Phi_{\pm}$ in Eq.~\eqref{eq:T-FLC})
and orientation of the optical axes
(the azimuthal angle $\psi_{d}$ in Eq.~\eqref{eq:T-FLC})
depend on the applied electric field.

The results
of a comprehensive theoretical
treatment of this effect
performed in Ref.~\cite{Kiselev:pre:2:2014}
are reviewed in Ref.~\cite{Kiselev:jmolliq:2018}.
Similarly to Ref.~\cite{Kiselev:pre:2015},
where electro-optic response
of DHFLC was studied for unpolarized light,
we have used these results to
fit the experimental data
(see Fig.~\ref{fig:expt_dhf})
and to evaluate 
electric field dependencies of
the geometric phases
using the parameters obtained
from the fitting procedure
(see Figs.~\ref{fig:phi_I_dhf}--\ref{fig:phi_U_016_dhf}).

Referring to Fig.~\ref{fig:phi_I_dhf},
the interferometric
phase appears to
be close to 
the Pancharatnam phase
when the polarized part
of the input wave
is linearly polarized.
From a comparison between
the Uhlmann and interferometric phases
plotted in Fig.~\ref{fig:phi_U_dhf},
it can be inferred that
a considerable difference between
the phases at relatively low degrees of polarization
vanishes  in the limit of pure state
(fully polarized wave) 
with $P_0\to 1$.
Figures~\ref{fig:phi_I_016_dhf} and~\ref{fig:phi_U_016_dhf}
demonstrate sensitivity
of the interferometric phase
to the ellipticity of the incident light
as compared to the Pancharatnam
and the Uhlmann phases.

Our concluding remark concerns
the interferometric phase in the regime
of nonunitary dynamics.
From formula~\eqref{eq:phi-psi},
the dynamical phases
that determine the interferometric
holonomies can, similar to the case of unitary evolution,
be expressed solely in terms of the
eigenpolarization states of the density matrix.
Analytical results presented in Appendix~\ref{sec:appendix2}
describe how the phases of the electrically controlled
birefringent plate are affected by the 
dissipation (nonunitarity) effects such as absorption
or Fresnel reflections.
A more comprehensive analysis and 
experimental studies of such effects
are now in progress.

\begin{acknowledgments} 
ADK acknowledges 
financial support from
the Ministry of Education and Science of the Russian Federation 
(project No.~RFMEFI58316X0058).

% This work is supported by 
% the
% RFBR grants 16-02-00441~A, 16-29-14012~ofi\_m
% and 16-42-630773~p\_a.
\end{acknowledgments}

%\bibliography{optics,polymer,scatter,lc,quant,hk,flc,qft,math,my_papers}


\appendix

%%%%%%%%%%%%%%
\section{Uhlmann connection}
\label{sec:appendix}
%%%%%%%%%%%%%%

In this Appendix we present details
on computing Uhlmann connection
$\vc{H}_U$ for the unitary
operators of the form:
\begin{align}
  \label{eq:T-general}
  \vc{T}=
\e^{i\eta (\uvc{m}\cdot\bs{\sigma})}=
\cos\eta\,\bs{\sigma}_0+i\sin\eta\,\sca{\uvc{m}}{\bs{\sigma}},
\end{align}
where $\uvc{m}=\uvc{m}(\psi)$
and the angles $\eta$ and $\psi$
 are generally functions of the governing parameter.
Such operators represent birefringent plates
and we  can use the algebraic identity
\begin{align}
  \label{eq:rel-3}
  \sca{\uvc{s}_0}{\bs{\sigma}} \sca{\uvc{m}}{\bs{\sigma}}=
  \sca{\uvc{s}_0}{\uvc{m}}\bs{\sigma}_0+
i
\sca{\uvc{s}_0\times\uvc{m}}{\bs{\sigma}}
\end{align}
to derive the density matrix
of light passed through such a plate
in the following form:
\begin{align}
&
  \label{eq:rho_eta}
  2\vc{T}\rho_0\hcnj{\vc{T}}\equiv 2\rho_\eta=\bs{\sigma}_0+
P_0 \sca{\uvc{s}_\eta}{\bs{\sigma}},
\\
&
\label{eq:mp}
\uvc{s}_\eta=\cos(2\eta)\uvc{s}_0+(1-\cos(2\eta))\sca{\uvc{s}_0}{\uvc{m}}\uvc{m}+
\notag
\\
&
\sin(2\eta)
\uvc{s}_0\times\uvc{m}.
\end{align}

We can also use the identity~\eqref{eq:rel-3}
to calculate the operator $\vc{H}$
\begin{align}
&
  \label{eq:H_0}
  \hcnj{\vc{T}}\dot{\vc{T}}=i\vc{H},\quad
  \vc{H}=\sca{\vc{q}}{\bs{\sigma}},\quad
\vc{q}=\dot{\eta}\,\uvc{m}+\dot{\psi}\sin\eta\,\vc{n},
  \\
&
\vc{n}=\cos\eta\,\uvc{m}'_{\psi}+\sin\eta\,\uvc{m}\times\uvc{m}'_{\psi},
\end{align}
where $\uvc{m}'_{\psi}=\prt{\psi}\uvc{m}$
and dot over the letter denotes the derivative with respect to
the governing parameter,
that enter equation~\eqref{eq:H_U-1} 
for the Uhlmann connection.

The relation~\eqref{eq:H_0}
can be generalized to
the case of the Wigner operators
$\vc{D}(\alpha,\beta,\gamma)$
[see Eq.~\eqref{eq:D_Wigner}]
with the angles being a function of
the governing parameter.
In this case, we have
\begin{align}
  \label{eq:H_D}
&
  \hcnj{\vc{D}}\dot{\vc{D}}=-i\vc{H}_D,\quad
  \vc{H}_D=\sca{\vc{d}}{\bs{\sigma}},
\\
&
  \label{eq:h_D}
d_1=-\dot{\alpha}\sin(2\beta)\cos(2\gamma)+\dot{\beta}\sin(2\gamma),
\notag
\\
&
d_2=\dot{\beta}\cos(2\gamma),\:
d_3=\dot{\alpha}\cos(2\beta)+\dot{\gamma}.
\end{align}
These results can be obtained using 
the algebraic identities
for the $1/2$-spin Wigner matrices~\cite{Varshalovich:bk:1988}
written in the following form:
\begin{align}
  \label{eq:rel-Pauli-D1}
&
  \mvc{D}\bs{\sigma}_i\hcnj{\mvc{D}}=
\sca{\uvc{m}_i}{\bs{\sigma}},
\\
  \label{eq:rel-Pauli-D2}
&
\hcnj{\mvc{D}}
\bs{\sigma}
\mvc{D}=
\uvc{m}_1\bs{\sigma}_1+
\uvc{m}_2\bs{\sigma}_2+
\uvc{m}_3\bs{\sigma}_3,
\end{align}
where 
$\uvc{m}_1=\cos(2\gamma)\uvc{s}_1+
\sin(2\gamma)\uvc{s}_2$,
$\uvc{m}_2=-\sin(2\gamma)\uvc{s}_1+
\cos(2\gamma)\uvc{s}_2$,
and  $\uvc{m}_3=\uvc{s}_3$;
the columns of the rotation matrix
\begin{align}
  \label{eq:rel-S}
&
\vc{S}=
  \begin{pmatrix}
    \cos(2\beta)\cos(2\alpha) & -\sin(2\alpha)&\sin(2\beta)\cos(2\alpha)\\
\cos(2\beta)\sin(2\alpha)&\cos(2\alpha)&\sin(2\beta)\sin(2\alpha)\\
-\sin(2\beta)&0&\cos(2\beta)
  \end{pmatrix}
\end{align}
give the components of $\uvc{s}_i$:
$\uvc{s}_i=\sum_k\vc{S}_{ki}\uvc{e}_k$.

For unitary evolution 
with $\vc{A}=\vc{I}$
and $\tilde{\rho}=\rho_0$,
Eq.~\eqref{eq:H_U-1} 
assumes the simplified form:
\begin{align}
  \label{eq:HU_H_uni}
  2\sqrt{\rho_0}\vc{H}\sqrt{\rho_0}=-[\vc{H}_U,\rho_0]_{+}.
\end{align}
In the basis of eigenstates of the density matrix $\rho_0$
we have
\begin{align}
&
  \label{eq:H_rho_0}
  \hcnj{\vc{C}}_0\vc{H}\vc{C}_0=
\sca{\vc{q}}{\uvc{s}_1}\bs{\sigma}_1+
\sca{\vc{q}}{\uvc{s}_2}\bs{\sigma}_2+
\sca{\vc{q}}{\uvc{s}_0}\bs{\sigma}_3
\\
&
  \label{eq:HU_rho_0}
  \hcnj{\vc{C}}_0\vc{H}_U\vc{C}_0=
\tilde{q}_1 \bs{\sigma}_1+
\tilde{q}_2 \bs{\sigma}_2+
\tilde{q}_3 \bs{\sigma}_3,
\\
&
  \label{eq:HHU_rho_0}
(1-P_0^2)^{1/2}
(\sca{\vc{q}}{\uvc{s}_1}\bs{\sigma}_1+
\sca{\vc{q}}{\uvc{s}_2}\bs{\sigma}_2)+
2\sca{\vc{q}}{\uvc{s}_0}\bs{\sigma}_3\rho_0=
\notag
\\
&
-(\tilde{q}_1 \bs{\sigma}_1+
\tilde{q}_2 \bs{\sigma}_2)
-2\tilde{q}_3 \bs{\sigma}_3\rho_0,
\end{align}
where 
$2\bs{\sigma}_3\rho_0=P_0\bs{\sigma}_0+\bs{\sigma}_3$
and
equation~\eqref{eq:HHU_rho_0}
is derived from Eq.~\eqref{eq:HU_H_uni}.
It is not difficult to find the components
of the operator~\eqref{eq:HU_rho_0}:
$\tilde{q}_{1,2}=-(1-P_0^2)^{1/2}\sca{\vc{h}}{\uvc{s}_{1,2}}$
and $\tilde{q}_3=-\sca{\vc{h}}{\uvc{s}_0}$.
Turning back to the circular basis
can be performed by replacing
$\bs{\sigma}_i$ with $\sca{\uvc{s}_i}{\bs{\sigma}}$.
The resulting expression for the Uhlmann connection reads
\begin{align}
&
  \label{eq:HU}
\vc{H}_U=
\sca{\vc{q}_U}{\bs{\sigma}}=
  \tilde{q}_1 \sca{\uvc{s}_1}{\bs{\sigma}}+
  \tilde{q}_2 \sca{\uvc{s}_2}{\bs{\sigma}}+
  \tilde{q}_2 \sca{\uvc{s}_1}{\bs{\sigma}},
\\
&
  \label{eq:hU}
\vc{q}_U=
-(1-P_0^2)^{1/2}\vc{q}-
[1-(1-P_0^2)^{1/2}]\sca{\vc{q}}{\uvc{s}_0}\uvc{s}_0.
\end{align}
Note that, for operators of the
form: $\vc{B}=b_0\bs{\sigma}_0+\sca{\vc{b}}{\bs{\sigma}}$,
considerations along similar lines
lead to the following relation:
\begin{align}
&
  \label{eq:rho_B_rho}
  2\sqrt{\rho_0}\vc{B}\sqrt{\rho_0}
=
(1-P_0^2)^{1/2}
\vc{B}+
\notag
\\
&
\bigl\{
[1-({1-P_0^2})^{1/2}]
\sca{\vc{b}}{\uvc{s}_0}
+P_0b_0
\bigr\}
\sca{\uvc{s}_0}{\bs{\sigma}}+
\notag
\\
&
\bigl\{
P_0\sca{\vc{b}}{\uvc{s}_0}+
[1-({1-P_0^2})^{1/2}]b_0
  \bigr\}
\bs{\sigma}_0.
\end{align}

%%%%%%%%%%%%%%
\section{Phases for birefringent plates and effects of dissipation}
\label{sec:appendix2}
%%%%%%%%%%%%%%

In this Appendix
we relax
the assumption of lossless
transmission and
extend our analysis
presented in Sec.~\ref{subsec:dhflc}
to a more general case 
of a birefringent plate characterized by the nonunitary
transmission matrix
\begin{align}
&
  \label{eq:T_abs}
  \vc{T}(\tau)=
\vc{C}_d
\vc{T}_d
\vc{U}_d
\hcnj{\vc{C}}_d,
\notag
\\
&
\vc{T}_d=
\begin{pmatrix}
    t_{+} & 0\\
    0 &t_{-} 
  \end{pmatrix},
\:
\vc{U}_d=
\begin{pmatrix}
    \e^{i\Phi_{+}} & 0\\
    0 &\e^{i\Phi_{-}} 
  \end{pmatrix},
\end{align}
where
$\vc{C}_d=\vc{C}(\pi/4,\psi_d)$
and
$0\le t_{\pm}\le 1$
are the transmittance coefficients
that take into account
the effects of losses such as
Fresnel reflections or absorption.
In the limiting case of unitary (lossless) evolution
analyzed in Sec.~\ref{subsec:dhflc}
these coefficients are both equal to unity:
$t_{\pm}=1$
and we obtain the transmission matrix
of the DHFLC cell given by Eq.~\eqref{eq:T-FLC}.

%%%%%%%%%%%%%%%%
\subsection{Interferometric phase}
\label{subsec:int-phase-app2}
%%%%%%%%%%%%%%%%

We begin with the interferometric phase.
Computing the interferometric
phase involves three steps:
(a)~solving the spectral problem
for the density matrix
$\rho(\tau)=\vc{T}(\tau)\rho_0\hcnj{\vc{T}}(\tau)$ to find
the eigenpolarization vectors 
$\ket{p_\mu(\tau)}$
and eigenvalues
$p_\mu(\tau)$;
(b)~evaluating
the dynamical phase~\eqref{eq:phi-psi};
and
(c)~substituting the unitary operator~\eqref{eq:tV_I}
into the expression for the interferometric
function~\eqref{eq:FI_rel}.

Our starting point is the expression for the density matrix
given by
\begin{align}
&
  \label{eq:rho_abs}
  \frac{2}{r_{+}}\hcnj{\vc{V}}_d\,\rho(\tau)\,\vc{V}_d=
q_{0}^{(\rho)}\bs{\sigma}_0+
\sca{\vc{q}_\rho}{\bs{\sigma}},
\\
&
  \label{eq:q_rho_abs}
q_{0}^{(\rho)}=1+r P_0 \sca{\uvc{s}_0}{\uvc{s}_d},\:
q_{3}^{(\rho)}=r+P_0
\sca{\uvc{s}_0}{\uvc{s}_{d}},
\notag
\\
&
q_{1,2}^{(\rho)}=P_0\sqrt{1-r^2}
\sca{\uvc{s}_0}{\uvc{s}_{1,2}^{(d)}},
\\
&
\label{eq:r_abs}
r=\frac{r_{-}}{r_{+}}=\frac{t_{+}^2-t_{-}^2}{t_{+}^2+t_{-}^2},
\end{align}
where
$\vc{V}_d=\vc{C}_d\vc{U}_d$
and 
$r_{\pm}=(t_{+}^2\pm t_{-}^2)/2$;
$\uvc{s}_{1}^{(d)}=(0,0,-1)$
and $\uvc{s}_{2}^{(d)}=
(-\sin(2\psi_d),\cos(2\psi_d),0)$.
We can now use
the identities~\eqref{eq:rel-Pauli-1} for 
the $1/2$-spin Wigner rotation martrices
to diagonalize the matrix~\eqref{eq:rho_abs}
as follows
\begin{align}
&
  \label{eq:C_l_abs}
  \hcnj{\vc{C}}_\rho\sca{\vc{q}_\rho}{\bs{\sigma}}\vc{C}_\rho=
|\vc{q}_\rho| \bs{\sigma}_3,\:
\vc{C}_\rho\equiv\vc{C}(\phi_\rho,\theta_\rho),
\end{align}
where
\begin{align}
&
  \label{eq:cs_theta_abs}
\cos(2\theta_\rho)=
\frac{r+P_0
  \sca{\uvc{s}_0}{\uvc{s}_{d}}}{|\vc{q}_\rho|},
\\
&
  \label{eq:abs_q_rho}
|\vc{q}_\rho|^2=
P_0^2(1-r^2)
\bigl[1-\sca{\uvc{s}_0}{\uvc{s}_{d}}^2
\bigr]
\notag
\\
&
+
\bigl[
r+P_0
  \sca{\uvc{s}_0}{\uvc{s}_{d}}
\bigr]^2.  
\end{align}
At this stage, we have 
the matrix of eigenpolarization vectors
\begin{align}
\label{eq:V_rho_abs}
\vc{V}_\rho=\vc{V}_d\vc{C}_\rho  
\end{align}
and the eigenvalues of the density matrix
\begin{align}
  \label{eq:p_mu_abs}
  p_\mu(\tau)=\frac{r_{+}}{2}
\bigl(
q_{0}^{(\rho)}+\mu |\vc{q}_\rho|
\bigr)\equiv 
\frac{\Tr\rho}{2}
(1+\mu\, P),
\end{align}
where $P=|\vc{q}_\rho|/q_{0}^{(\rho)}$
is the degree of polarization.
Note that,
in Eqs.~\eqref{eq:q_rho_abs}--~\eqref{eq:V_rho_abs},
the transmittance anisotropy
parameter $r$ [see Eq.~\eqref{eq:r_abs}]
is the only parameter
describing
nonunitarity effects.
These effects also result
in reduction of
the transmitted light  intensity,
so that the trace of the density matrix 
$\Tr\rho(\tau)=r_{+}q_{0}^{(\rho)}$ is generally differ from
unity.

For the eigenpolarization vectors~\eqref{eq:V_rho_abs}
and the transmission matrix~\eqref{eq:T_abs},
it is not difficult to obtain 
the dynamical phases
 \begin{align}
&
   \label{eq:Phi_d_abs}
   \Phi_{\mu}^{(d)}=
\Im 
\int_{0}^{h}
\bra{p_{\mu}}\dot{\vc{T}}\vc{T}^{-1}\ket{p_{\mu}}\dd\tau=\Phi+\mu{\Phi}_d,
\notag
\\
&
{\Phi}_d=\int_{0}^{h}\cos(2\theta_\rho)\Delta\dot{\Phi}(\tau)\dd\tau,
 \end{align}
giving the interferometric function
 \begin{align}
&
   \label{eq:F_I_abs}
   F_I=\Tr[\rho_0\tilde{\vc{V}}_I(h)\vc{T}(h)]
\notag
\\
&
=
\Tr\left\{
\tilde{\vc{V}}_I(h)\rho(h)[\hcnj{\vc{T}}(h)]^{-1}
\right\}
\notag
\\
&
=\sum_{\mu=\pm 1}
\e^{-i\Phi_\mu^{(d)}}
p_\mu
\bra{p_\mu}[\hcnj{\vc{T}}]^{-1}
\ket{p_\mu}
\notag
\\
&
=\frac{\Tr\rho}{2}\sum_{\mu=\pm 1}
\e^{-i\mu{\Phi}_d}
(1+\mu\, P)
[t_0+\mu\, t_3\cos(2\theta_\rho)],
 \end{align}
where
 \begin{align}
   \label{eq:t0_t3}
   \vc{T}_d^{-1}
\vc{U}_d=\e^{i\Phi}
(t_0\bs{\sigma}_0+t_3\bs{\sigma}_3),
 \end{align}
that defines the interferometric
phase: $\Phi_I=\arg F_I$.
It can be checked that,
in the unitary limit with $r=0$,
formula~\eqref{eq:F_I_abs}
reproduces
the result
given by Eq.~\eqref{eq:F_I_flc}
of Sec.~\ref{subsec:dhflc}. 

%%%%%%%%%%%%%%%%
\subsection{Uhlmann phase}
\label{subsec:uhl-phase-app2}
%%%%%%%%%%%%%%%%

Now we turn to 
the Uhlmann phase
and start with computing
the Uhlmann connection.
This connection can be found
by solving Eq.~\eqref{eq:H_U-1}.
After evaluating
the operators that enter
both sides of this equation,
we find
that, similar to Eq.~\eqref{eq:HU_H_uni},
it can be written in the form:
\begin{align}
  \label{eq:Eq-HU_abs}
  2\vc{H}=-[\vc{H}_U,\tilde{\rho}]_{+},
\end{align}
where
\begin{align}
&
  \label{eq:H_abs}
  \vc{H}=\sqrt{\rho_0}\vc{H}_0\sqrt{\rho_0},
\:
\vc{H}_0=
\vc{C}_d
\begin{pmatrix}
    t_{+}^2\dot{\psi}_{+} & 0\\
    0 &t_{-}^2\dot{\psi}_{-} 
  \end{pmatrix}
\hcnj{\vc{C}}_d,
\\
&
\label{eq:trho_abs}
  \tilde{\rho}=\sqrt{\rho_0}\vc{A}\sqrt{\rho_0},
\quad
\vc{A}=
\vc{C}_d
\begin{pmatrix}
    t_{+}^2 & 0\\
    0 &t_{-}^2 
  \end{pmatrix}
\hcnj{\vc{C}}_d.
\end{align}
We can now substitute the operators
\begin{align}
&
  \label{eq:H_exp_abs}
  \vc{H}=
h_0\bs{\sigma}_0+\sca{\vc{q}_0}{\bs{\sigma}},
\quad
  \tilde{\rho}=
h_A\bs{\sigma}_0+\sca{\vc{q}_A}{\bs{\sigma}},
\notag
\\
&
\vc{H}_U=
-h_U\bs{\sigma}_0-\sca{\vc{q}_U}{\bs{\sigma}}
\end{align}
into Eq.~\eqref{eq:Eq-HU_abs}
to derive a system of linear equations
\begin{align}
&
  \label{eq:syst_abs}
  \begin{cases}
  h_U h_A +\sca{\vc{q}_U}{\vc{q}_A}=h_0\\
h_U\vc{q}_A+h_A\vc{q}_U=\vc{q}_0  
  \end{cases}
\end{align}
with the solution given by
\begin{subequations}
  \label{eq:sols_abs}
\begin{align}
&
  \label{eq:hU_abs}
  h_U=\frac{h_0
  h_A-\sca{\vc{q}_U}{\vc{q}_A}^2}{h_A^2-|\vc{q}_A|^2}=\dot{\Phi},
\\
&
  \label{eq:qU_abs}
\vc{q}_U=(\vc{q}_0-h_U\vc{q}_A)/h_A.  
\end{align}  
\end{subequations}
The resulting expression for the Uhlmann connection
reads
\begin{align}
&
  \label{eq:HU_abs}
  \vc{H}_U=-\dot{\Phi}\bs{\sigma}_0-
\frac{\Delta\dot{\Phi}\sca{\vc{k}_U}{\bs{\sigma}}}
{1+r P_0 \sca{\uvc{s}_0}{\uvc{s}_d}}
\notag
\\
&
=-\dot{\Phi}\bs{\sigma}_0-
\Delta\dot{\Phi}\gamma\sca{\uvc{k}_U}{\bs{\sigma}},
\end{align}
where
\begin{align}
&
\label{eq:k_u_abs}
\vc{k}_U=(1-P_0^2)^{1/2}(\uvc{s}_d-\sca{\uvc{s}_0}{\uvc{s}_{d}}\uvc{s}_0)
\notag
\\
&
+
\left\{
rP_0+
\sca{\uvc{s}_0}{\uvc{s}_{d}}
\right\}
\uvc{s}_0,
\\
&
  \label{eq:gamma_abs}
\gamma^2=
1-\frac{P_0^2(1-r^2)[1-\sca{\uvc{s}_0}{\uvc{s}_{d}}^2]}{[1+rP_0 
\sca{\uvc{s}_0}{\uvc{s}_{d}}]^2}.
\end{align}
In contrast to the case of unitary evolution
described by Eqs.~\eqref{eq:H_U_flc}--\eqref{eq:F_U_flc},
the connection~\eqref{eq:HU_abs}
generally depends on the governing (thickness)
parameter $\tau$
as the transmittance anisotropy parameter $r$
is a function of $\tau$.
So, the Uhlmann holonomy
for the above connection
is given by
\begin{align}
  \label{eq:V_U_abs}
  \vc{V}_U(h)=\e^{-i\Phi}\mathcal{T}\e^{-i\int_0^h\Delta\dot{\Phi}\gamma\sca{\uvc{k}_U}{\bs{\sigma}}\dd\tau},
\end{align}
where $\mathcal{T}$ is 
the evolution parameter ordering
operator along the path,
and we obtain the Uhlmann phase
\begin{align}
  \label{eq:F_U_abs}
\Phi_U=\arg F_U,\:
  F_U=\Tr[\sqrt{\rho_0}\vc{T}(h)\sqrt{\rho_0}\vc{V}_U(h)]
\end{align}
expressed in terms of the Uhlmann function.

\end{document}